\documentclass{iopjournal}
\usepackage{amssymb} 
\usepackage{graphicx}
\usepackage{bbm}
\usepackage{bm}
\usepackage{amsmath}
\usepackage{mathtools}
\usepackage[dvipsnames]{xcolor}
\usepackage{hyperref}
\usepackage{tikz}
\usepackage{CJK}
\usepackage{url}
\usepackage{lmodern}

\definecolor{darkblue}{rgb}{0,0,0.6}

\hypersetup{
bookmarksopen=true,
pdftitle="",
pdfauthor="", 
pdftoolbar=false,
pdfstartview={FitH},
pdfmenubar=true,
pdfhighlight=/O,
colorlinks=true,
urlcolor=darkblue,
citecolor=darkblue,
linkcolor=darkblue}

\usetikzlibrary{decorations.pathmorphing}
\usetikzlibrary{decorations.markings}
\usetikzlibrary{decorations}
\pgfkeys{
  /tikz/segment amplitude/.style={/tikz/decoration/amplitude=#1},
  /tikz/line after coil/.style={/tikz/decoration/post length=#1},
  /tikz/line before coil/.style={/tikz/decoration/pre length=#1},
  /tikz/segment length/.style={/tikz/decoration/segment length=#1},
  /tikz/segment aspect/.style={/tikz/decoration/aspect=#1},
}
\colorlet{DiagColor}{darkblue}
\tikzset{
xxtsubstrate/.style={decorate, 
line width=1pt,
draw=olive, 
decoration=coil, 
segment amplitude=0.75mm, 
line after coil=0.25mm,
line before coil=0.25mm
},
tsubstrate/.style={decorate, 
line width=1pt,
draw=olive, 
decoration=coil, 
segment amplitude=0.5mm, 
segment length=5pt,
segment amplitude=0.2mm, 
line after coil=1mm,
line before coil=1mm
},
Bsubstrate/.style={decorate, 
line width=1pt,
draw=olive, 
decoration=coil,
segment length=5pt,
segment aspect=0,
segment amplitude=0.5mm, 
line after coil=0mm,
line before coil=0mm
},
substrate/.style={decorate, 
line width=1pt,
draw=olive, 
decoration=coil, 
segment length=5pt,
segment amplitude=0.5mm, 
line after coil=0.5mm,
line before coil=0.5mm
},
activity/.style={very thick,draw=DiagColor,postaction={decorate},
decoration={markings,mark=at position .5 with
{\arrow[draw=DiagColor]{>}}}},
tactivity/.style={thick,draw=DiagColor,postaction={decorate},
decoration={markings,mark=at position .5 with
{\arrow[draw=DiagColor]{>}}}},
tEPSactivity/.style={thick,draw=DiagColor,postaction={decorate},
decoration={markings,mark=at position .55 with
{\arrow[draw=DiagColor]{>}}}},
tAactivity/.style={thick,draw=DiagColor},
Aactivity/.style={very thick,draw=DiagColor},
ghostactivity/.style={very thick,draw=white},
Bactivity/.style={very thick,draw=blue,dashed},
Baractivity/.style={very thick,draw=black},
Cactivity/.style={very thick,draw=cyan,decoration=coil},
tSactivity/.style={thick,draw=DiagColor,postaction={decorate},
decoration={markings,mark=at position .7 with
{\arrow[draw=DiagColor]{>}}}},
Sactivity/.style={very thick,draw=DiagColor,postaction={decorate},
decoration={markings,mark=at position .7 with
{\arrow[draw=DiagColor]{>}}}},
ABPactivity/.style={very thick,draw=DiagColor,decorate,decoration={coil,segment length=6pt}},
Arrowactivity/.style={decoration={markings,mark=at position 1 with
    {\arrow[scale=1.6,>=stealth]{>}}},postaction={decorate}},
triangle/.style = {fill=blue!10, regular polygon, regular polygon sides=3,minimum size=3mm},
    node rotated/.style = {rotate=180},
    border rotated/.style = {shape border rotate=180},
    diamonds/.style = {fill=blue!10,inner sep=0pt,minimum
size=0.2cm,inner sep=0pt,outer sep=0pt},}

\tikzset{coil it/.style={decorate, decoration={coil, segment length=1.5mm, amplitude=0.25mm}},Tactivity/.style={very thick,draw=DiagColor,coil it},}

\newcommand{\velpertXa}[2]{
\tikz[baseline=-1pt]{
\draw[Aactivity]  (0,0)--(180:0.6)  node[at end ,above]{$#1$};
\draw[Aactivity]  (0,0)--(0:0.6)  node[at end ,below=2pt]{$#2$};
\draw[Baractivity]  (-0.2,-0.1)--(-0.2,0.1);
    \node[isosceles triangle, isosceles triangle apex angle=60,
    draw,
    rotate=90,
    fill=DiagColor,color=DiagColor,scale=0.5] (b) at (0,0){};
}}

\newcommand{\velpertXb}[2]{
\tikz[baseline=-1pt]{
\draw[Aactivity]  (0,0)--(180:0.6)  node[at end ,above]{$#1$};
\draw[Aactivity]  (0,0)--(0:0.6)  node[at end ,below=2pt]{$#2$};
\draw[Baractivity]  (-0.2,-0.1)--(-0.2,0.1);
    \node[isosceles triangle, isosceles triangle apex angle=60,
    draw,
    rotate=270,
    fill=DiagColor,color=DiagColor,scale=0.5] (b) at (0,0){};
}}

\newcommand{\tumblepert}[2]{
\tikz[baseline=-1pt]{
\draw[Aactivity]  (0,0)--(180:0.6)  node[at end ,above]{$#1$};
\draw[Aactivity]  (0,0)--(0:0.6)  node[at end ,below=2pt]{$#2$};
\draw[Aactivity] (-0,0) -- (-0,0) node[diamonds,draw=black]{};
}}

\newcommand{\barepropX}[2]{\tikz[baseline=-2.5pt]{
\draw[Aactivity] (180:1.2) -- (0,0)  node[at start ,above]{$#1$} ;
\draw[Aactivity] (0:0.6) -- (0,0) node[at start,below=2pt] {$#2$};}}

\newcommand{\upvert}[2]{
\tikz[baseline=-1pt]{
\draw[Aactivity] (0,0)--(180:0.6)  node[at end ,above]{$#1$};
\draw[Aactivity] (0,0)--(0:0.6)  node[at end ,below=2pt]{$#2$};
\node[isosceles triangle, isosceles triangle apex angle=60,
draw,
rotate=90,
fill=DiagColor,color=DiagColor,scale=0.6] (b) at (0,0){};
}}

\newcommand{\downvert}[2]{
\tikz[baseline=-1pt]{
\draw[Aactivity]  (0,0)--(180:0.6)  node[at end ,above]{$#1$};
\draw[Aactivity]  (0,0)--(0:0.6)  node[at end ,below=2pt]{$#2$};
    \node[isosceles triangle, isosceles triangle apex angle=60,
    draw,
    rotate=270,
    fill=DiagColor,color=DiagColor,scale=0.6] (b) at (0,0){};
}}

\newcommand{\tvert}[2]{
\tikz[baseline=-1pt]{
\draw[Aactivity]  (0,0)--(180:0.6)  node[at end ,above]{$#1$};
\draw[Aactivity]  (0,0)--(0:0.6)  node[at end ,below=2pt]{$#2$};
\node[circle,
fill=DiagColor,color=DiagColor,scale=0.75] (b) at (0,0){};
}}

\newcommand{\fullprop}[2]{
\tikz[baseline=-1pt]{
\draw[Aactivity]  (0,0)--(180:0.75)  node[at end ,above]{$#1$};
\draw[Aactivity]  (0,0)--(0:0.75)  node[at end ,below=2pt]{$#2$};
\node[circle,
fill=white,draw=DiagColor,scale=1] (b) at (0,0){};
}}

\newcommand{\barepropcurlytwo}[2]{\tikz[baseline=-2.5pt]{
\draw[Tactivity] (180:1.2) -- (0,0)  node[at start ,above]{$#1$} node[at end,below=2pt] {$#2$};}}

\newcommand{\ttermone}[2]{
\tikz[baseline=-2.5pt]{
\draw[Aactivity]  (0,0)--(180:0.6)  node[at end ,above]{$#1$};
\draw[Aactivity]  (0,0)--(0:0.6)  node[at end ,below=2pt]{$#2$};
\node[circle,
fill=DiagColor,color=DiagColor,scale=0.75] (b) at (0,0){};
}}

\newcommand{\ttermtwo}[2]{
\tikz[baseline=-2.5pt]{
\draw[Aactivity]  (0,0)--(180:0.6)  node[at end ,above]{$#1$};
\draw[Aactivity]  (0,0)--(0:0.6);
\draw[Aactivity]  (0.6,0)--(0:1.2)  node[at end ,below=2pt]{$#2$};
\node[circle,
fill=DiagColor,color=DiagColor,scale=0.75] (b) at (0,0){};
\node[circle,
fill=DiagColor,color=DiagColor,scale=0.75] (b) at (0.6,0){};
}}

\newcommand{\barepropshorttwo}[2]{\tikz[baseline=-2.5pt]{
\draw[Aactivity] (180:0.3) -- (0,0)  node[at start ,above]{$#1$} ;
\draw[Aactivity] (0:0.3) -- (0,0) node[at start,below=2pt] {$#2$};}}

\newcommand{\barepropcurlyshort}[2]{\tikz[baseline=-2.5pt]{
\draw[Tactivity] (180:0.6) -- (0,0)  node[at start ,above]{$#1$} node[at end,below=2pt] {$#2$};}}

\newcommand{\upprop}[2]{\tikz[baseline=-2.5pt]{
\draw[Tactivity] (0:1.2) -- (0,0)  node[at start ,below=2pt]{$#1$} node[at end,above] {$#2$};
    \node[isosceles triangle, isosceles triangle apex angle=60,
    draw,
    rotate=90,
    fill=DiagColor,color=DiagColor,scale=0.5] (b) at (0.6,0){};}}

\newcommand{\downprop}[2]{\tikz[baseline=-2.5pt]{
\draw[Tactivity] (0:1.2) -- (0,0)  node[at start ,below=2pt]{$#1$} node[at end,above] {$#2$};
    \node[isosceles triangle, isosceles triangle apex angle=60,
    draw,
    rotate=270,
    fill=DiagColor,color=DiagColor,scale=0.5] (b) at (0.6,0){};}}

\newcommand{\downupprop}[2]{\tikz[baseline=-2.5pt]{
\draw[Tactivity] (180:0.6) -- (0,0)  node[at start ,above]{$#1$};
\draw[Tactivity] (0:0.6) -- (0,0);
\draw[Tactivity] (0.6,0) -- (1.2,0) node[at end,below=2pt] {$#2$};
    \node[isosceles triangle, isosceles triangle apex angle=60,
    draw,
    rotate=90,
    fill=DiagColor,color=DiagColor,scale=0.5] (b) at (0,0){};
    \node[isosceles triangle, isosceles triangle apex angle=60,
    draw,
    rotate=270,
    fill=DiagColor,color=DiagColor,scale=0.5] (b) at (0.6,0){};}}

\newcommand{\updownprop}[2]{\tikz[baseline=-2.5pt]{
\draw[Tactivity] (180:0.6) -- (0,0)  node[at start ,above]{$#1$};
\draw[Tactivity] (0:0.6) -- (0,0);
\draw[Tactivity] (0.6,0) -- (1.2,0) node[at end,below=2pt] {$#2$};
    \node[isosceles triangle, isosceles triangle apex angle=60,
    draw,
    rotate=270,
    fill=DiagColor,color=DiagColor,scale=0.5] (b) at (0,0){};
    \node[isosceles triangle, isosceles triangle apex angle=60,
    draw,
    rotate=90,
    fill=DiagColor,color=DiagColor,scale=0.5] (b) at (0.6,0){};}}

\newcommand{\fourthpropa}[2]{\tikz[baseline=-2.5pt]{
\draw[Tactivity] (180:0.6) -- (0,0)  node[at start ,above]{$#1$};
\draw[Tactivity] (0:0.6) -- (0,0);
\draw[Tactivity] (0.6,0) -- (1.2,0);
\draw[Tactivity] (1.2,0) -- (1.8,0);
\draw[Tactivity] (1.8,0) -- (2.4,0) node[at end,below=2pt] {$#2$};
    \node[isosceles triangle, isosceles triangle apex angle=60,
    draw,
    rotate=270,
    fill=DiagColor,color=DiagColor,scale=0.5] (b) at (0,0){};
    \node[isosceles triangle, isosceles triangle apex angle=60,
    draw,
    rotate=90,
    fill=DiagColor,color=DiagColor,scale=0.5] (b) at (0.6,0){};
        \node[isosceles triangle, isosceles triangle apex angle=60,
    draw,
    rotate=270,
    fill=DiagColor,color=DiagColor,scale=0.5] (b) at (1.2,0){};
    \node[isosceles triangle, isosceles triangle apex angle=60,
    draw,
    rotate=90,
    fill=DiagColor,color=DiagColor,scale=0.5] (b) at (1.8,0){};}}

\newcommand{\fourthpropb}[2]{\tikz[baseline=-2.5pt]{
\draw[Tactivity] (180:0.6) -- (0,0)  node[at start ,above]{$#1$};
\draw[Tactivity] (0:0.6) -- (0,0);
\draw[Tactivity] (0.6,0) -- (1.2,0);
\draw[Tactivity] (1.2,0) -- (1.8,0); 
\draw[Tactivity] (1.8,0) -- (2.4,0) node[at end,below=2pt] {$#2$};
    \node[isosceles triangle, isosceles triangle apex angle=60,
    draw,
    rotate=90,
    fill=DiagColor,color=DiagColor,scale=0.5] (b) at (0,0){};
    \node[isosceles triangle, isosceles triangle apex angle=60,
    draw,
    rotate=270,
    fill=DiagColor,color=DiagColor,scale=0.5] (b) at (0.6,0){};
        \node[isosceles triangle, isosceles triangle apex angle=60,
    draw,
    rotate=90,
    fill=DiagColor,color=DiagColor,scale=0.5] (b) at (1.2,0){};
    \node[isosceles triangle, isosceles triangle apex angle=60,
    draw,
    rotate=270,
    fill=DiagColor,color=DiagColor,scale=0.5] (b) at (1.8,0){};}}

\newcommand{\fourthpropc}[2]{\tikz[baseline=-2.5pt]{
\draw[Tactivity] (180:0.6) -- (0,0)  node[at start ,above]{$#1$};
\draw[Tactivity] (0:0.6) -- (0,0);
\draw[Tactivity] (0.6,0) -- (1.2,0);
\draw[Tactivity] (1.2,0) -- (1.8,0);
\draw[Tactivity] (1.8,0) -- (2.4,0) node[at end,below=2pt] {$#2$};
    \node[isosceles triangle, isosceles triangle apex angle=60,
    draw,
    rotate=270,
    fill=DiagColor,color=DiagColor,scale=0.5] (b) at (0,0){};
    \node[isosceles triangle, isosceles triangle apex angle=60,
    draw,
    rotate=270,
    fill=DiagColor,color=DiagColor,scale=0.5] (b) at (0.6,0){};
        \node[isosceles triangle, isosceles triangle apex angle=60,
    draw,
    rotate=90,
    fill=DiagColor,color=DiagColor,scale=0.5] (b) at (1.2,0){};
    \node[isosceles triangle, isosceles triangle apex angle=60,
    draw,
    rotate=90,
    fill=DiagColor,color=DiagColor,scale=0.5] (b) at (1.8,0){};}}

\newcommand{\fourthpropd}[2]{\tikz[baseline=-2.5pt]{
\draw[Tactivity] (180:0.6) -- (0,0)  node[at start ,above]{$#1$};
\draw[Tactivity] (0:0.6) -- (0,0);
\draw[Tactivity] (0.6,0) -- (1.2,0);
\draw[Tactivity] (1.2,0) -- (1.8,0); 
\draw[Tactivity] (1.8,0) -- (2.4,0) node[at end,below=2pt] {$#2$};
    \node[isosceles triangle, isosceles triangle apex angle=60,
    draw,
    rotate=90,
    fill=DiagColor,color=DiagColor,scale=0.5] (b) at (0,0){};
    \node[isosceles triangle, isosceles triangle apex angle=60,
    draw,
    rotate=90,
    fill=DiagColor,color=DiagColor,scale=0.5] (b) at (0.6,0){};
        \node[isosceles triangle, isosceles triangle apex angle=60,
    draw,
    rotate=270,
    fill=DiagColor,color=DiagColor,scale=0.5] (b) at (1.2,0){};
    \node[isosceles triangle, isosceles triangle apex angle=60,
    draw,
    rotate=270,
    fill=DiagColor,color=DiagColor,scale=0.5] (b) at (1.8,0){};}}

\newcommand{\fourthprope}[2]{\tikz[baseline=-2.5pt]{
\draw[Tactivity] (180:0.6) -- (0,0)  node[at start ,above]{$#1$};
\draw[Tactivity] (0:0.6) -- (0,0);
\draw[Tactivity] (0.6,0) -- (1.2,0);
\draw[Tactivity] (1.2,0) -- (1.8,0); 
\draw[Tactivity] (1.8,0) -- (2.4,0) node[at end,below=2pt] {$#2$};
    \node[isosceles triangle, isosceles triangle apex angle=60,
    draw,
    rotate=270,
    fill=DiagColor,color=DiagColor,scale=0.5] (b) at (0,0){};
    \node[isosceles triangle, isosceles triangle apex angle=60,
    draw,
    rotate=90,
    fill=DiagColor,color=DiagColor,scale=0.5] (b) at (0.6,0){};
        \node[isosceles triangle, isosceles triangle apex angle=60,
    draw,
    rotate=90,
    fill=DiagColor,color=DiagColor,scale=0.5] (b) at (1.2,0){};
    \node[isosceles triangle, isosceles triangle apex angle=60,
    draw,
    rotate=270,
    fill=DiagColor,color=DiagColor,scale=0.5] (b) at (1.8,0){};}}

 \newcommand{\fourthpropf}[2]{\tikz[baseline=-2.5pt]{
\draw[Tactivity] (180:0.6) -- (0,0)  node[at start ,above]{$#1$};
\draw[Tactivity] (0:0.6) -- (0,0);
\draw[Tactivity] (0.6,0) -- (1.2,0);
\draw[Tactivity] (1.2,0) -- (1.8,0); 
\draw[Tactivity] (1.8,0) -- (2.4,0) node[at end,below=2pt] {$#2$};
    \node[isosceles triangle, isosceles triangle apex angle=60,
    draw,
    rotate=90,
    fill=DiagColor,color=DiagColor,scale=0.5] (b) at (0,0){};
    \node[isosceles triangle, isosceles triangle apex angle=60,
    draw,
    rotate=270,
    fill=DiagColor,color=DiagColor,scale=0.5] (b) at (0.6,0){};
        \node[isosceles triangle, isosceles triangle apex angle=60,
    draw,
    rotate=270,
    fill=DiagColor,color=DiagColor,scale=0.5] (b) at (1.2,0){};
    \node[isosceles triangle, isosceles triangle apex angle=60,
    draw,
    rotate=90,
    fill=DiagColor,color=DiagColor,scale=0.5] (b) at (1.8,0){};}}

\usetikzlibrary{decorations.pathmorphing}
\usetikzlibrary{decorations.markings}
\usetikzlibrary{shapes.geometric}

\newcommand{\latin}[1]{{\it #1}}
\newcommand{\ie}{\latin{i.e.}\ }

\newcommand{\plaind}{\mathrm{d}}
\newcommand{\imag}{\mathring{\imath}}
\renewcommand{\exp}[1]{\mathchoice{\mathrm{e}^{#1}}{\operatorname{exp}\left(#1\right)}{\operatorname{exp}\left(#1\right)}{\operatorname{exp}\left(#1\right)}}

\newcommand{\dbar}{\plaind\mkern-6mu\mathchar'26}
\newcommand{\dint}[1]{\mathchoice{\!\plaind#1\,}{\!\plaind#1\,}{\!\plaind#1\,}{\!\plaind#1\,}}
\newcommand{\ddintx}[2]{\mathchoice{\!\plaind^{#2}#1\,}{\!\plaind^{#2}#1\,}{\!\plaind^{#2}#1\,}{\!\plaind^{#2}#1\,}}
\newcommand{\ddint}[1]{\ddintx{#1}{d}}
\newcommand{\ddintbar}[1]{\mathchoice{\!\dbar^d#1\,}{\!\dbar^d#1\,}{\!\dbar^d#1\,}{\!\dbar^d#1\,}}

\newcommand{\gpvec}[1]{\mathbf{#1}}
\newcommand{\zerovec}{\gpvec{0}}
\newcommand{\vvec}{\gpvec{v}}
\newcommand{\nullvec}{\zerovec}
\newcommand{\dvec}{\gpvec{d}}
\newcommand{\kvec}{\gpvec{k}}
\newcommand{\mvec}{\gpvec{m}}
\newcommand{\xvec}{\gpvec{x}}
\newcommand{\transpose}{\mathsf{T}}

\newcommand{\AC}{\mathcal{A}}
\newcommand{\NC}{\mathcal{N}}
\newcommand{\action}{\AC}
\newcommand{\norm}{\NC}

\newcommand{\Ytilde}{\tilde{Y}}

\newcommand{\tumbleRate}{\alpha}
\newcommand{\ave}[1]{\left\langle #1 \right\rangle}

\newcommand{\elabel}[1]{\label{eq:#1}}
\newcommand{\eref}[1]{(\ref{eq:#1})}
\newcommand{\Eref}[1]{Eq.~(\ref{eq:#1})}
\newcommand{\Erefs}[1]{Eqs.~(\ref{eq:#1})}
\newcommand{\Sref}[1]{Section~\ref{sec:#1}}
\newcommand{\fref}[1]{Fig.~\ref{fig:#1}}

\newcommand{\effetumbleRate}{\Lambda}
\newcommand{\Gegenbauer}[1]{C^{(#1-2)/2}}
\newcommand{\Dx}{D}
\def\deltabar{{\mathchar '26\mkern -10mu\delta}}

\renewcommand{\Re}[1]{\mathfrak{R}(#1)}
\renewcommand{\Im}[1]{\mathfrak{I}(#1)}

\allowdisplaybreaks
\CJKnospace

\begin{document}
\articletype{Paper}

\title{Run-and-tumble particles with preferred reorientation}

\author{Callum Britton$^{1,\dagger}$, Ziluo Zhang$^{2,\dagger}$, Seongjun Han$^1$ and Thibault Bertrand$^{1,*}$}

\affil{$^1$ Department of Mathematics, Imperial College London, London SW7 2AZ, United Kingdom}

\affil{$^2$ Department of Physics, Xiamen University, Xiamen, Fujian 361005, China}

\affil{$^\dagger$ These authors contributed equally.}

\affil{$^*$ Author to whom correspondence should be addressed.}

\email{t.bertrand@imperial.ac.uk}

\keywords{active matter, run-and-tumble, field theory}

\begin{abstract}
Run-and-tumble particles (RTPs) are canonically modeled with uniform reorientation probabilities, an assumption that breaks down for many biological microswimmers. In this work, we investigate the dynamics of RTPs with arbitrary non-uniform tumble distributions. By deriving an exact Doi-Peliti field theory, we explicitly calculate a wide array of spatial and orientational observables. Notably, we demonstrate that the spatial dynamics exhibit an effective persistence and chirality governed entirely by the first Fourier modes of the tumble distribution, establishing a formal mapping to the dynamics of chiral active Brownian particles. Furthermore, our field-theoretic framework provides a systematic method to compute spatial moments to arbitrary order, allowing for the complete characterization and identification of complex tumbling dynamics. We illustrate the framework with wrapped Gaussian and bimodal Gaussian distributions, demonstrating explicit control over persistence and chirality. We further extend the field theory to $d$ dimensions, recovering the mean squared displacement in terms of a single effective tumble rate. Our results establish a direct link between the shape of the tumble distribution and the emergent dynamics, and provide a foundation for the study of interacting RTPs with non-uniform reorientation.
\end{abstract}


\section{Introduction}
\label{sec:intro}

Active matter---systems of agents that individually transduce energy from their surroundings into directed mechanical motion---has emerged as one of the most productive areas of nonequiibrium statistical mechanics over the past two decades \cite{Marchetti2013,Bechinger2016,Cates2012}. The richness of active matter phenomenology, from the collective migration of bacterial colonies to the flocking of birds \cite{Berg2004,Copeland2009,Elgeti2015,Vicsek1995}, has captured the imagination of many modern theoretical physicists which motivated the development of minimal models that isolate the essential physics of self-propulsion while remaining analytically tractable. 

Two such models have proven particularly influential: the active Brownian particle (ABP) \cite{Howse2007,Romanczuk2012,Fily2012} and the run-and-tumble particle (RTP) \cite{Tailleur2008,Cates2013}. In the RTP model, an example of a velocity-jump process \cite{VanKampen2007, Klages2008, DOrsogna2026, BressloffNewby2011}, a particle moves at constant speed $v$ along a fixed direction interrupted by sudden reorientation events, {\it tumbles}, that occur at a Poissonian rate $\alpha$. This model was originally motivated by the motility of {\it Escherichia coli}, which swims in nearly straight {\it runs} punctuated by rapid reorientations of its flagellar bundle \cite{Berg1972,Schnitzer1993,Berg2004}. In its standard (albeit simplistic) formulation, the post-tumble orientation is drawn uniformly from the interval $[-\pi,\pi]$, making the tumble angle distribution isotropic. Several studies of this simple model have characterized the dynamics arising from tumbling via a uniformly distributed angle \cite{Tailleur2008, Cates2013, Renadheer2019, Sevilla2019, Solon2015, Mallmin2019, Shreshtha2019, Basu2020, Garcia-Millan2021, Zhang2022a, Sezik2026}; for instance, it has been shown to produce long-time diffusive behaviour with effective diffusivity $D_\textrm{eff} = D + v^2/d\alpha$.

This assumption of uniform tumbling is, however, known to be biologically inaccurate. Seminal experiments by Berg and Brown \cite{Berg1972} already demonstrated that {\it E.\,coli} reorientations are biased: the bacterium preferentially continues in the roughly the same direction after a tumble, with a mean reorientation angle of approximately $68^\circ$ rather than $180^\circ$. Long timescale tracking of {\it E.\,coli} was made possibly by recent important experimental advances in 3D Lagrangian tracking confirming the nonuniformity of tumble reorientation dynamics \cite{Figueroa-Morales2020}. More broadly, microorganisms navigating chemical gradients, fluid flows, or complex and crowded environments may exhibit strongly anisotropic tumble angle distributions that encode information about their surroundings \cite{deGennes2004,Celani2010,Saragosti2012,Junot2022,Urra2025}. Understanding how the shape of the tumble distribution affects macroscopic transport properties is therefore not merely a theoretical exercise: it is a prerequisite for connecting minimal models to experimental observations. 

Despite this, the theoretical study of RTPs with non-uniform tumble distributions has remained limited. Existing analytical approaches have largely focused on specific distribution families or restricted themselves to calculating low order observables like the mean squared displacement and orientation autocorrelator \cite{Olsen2024}. A systematic framework capable of computing observables of arbitrary order, thereby fully characterizing the particle's dynamics, has been lacking. 

Field-theoretic methods have proven to be precisely the right tools for such systematic calculations in active matter. The Martin-Siggia-Rose (MSR) formalism \cite{MartinSiggiaRose1973,Janssen1976,deDominicis1976} and more recently, the Doi-Peliti (DP) field theory \cite{Doi1976a, Peliti1985} have been applied successfully to ABPs, RTPs, and their variants \cite{Bothe2021,Garcia-Millan2021,Zhang2022a,Pruessner2025,Roberts2022,Zhang2024,Britton2025,Sezik2026,Scandolo2023,Silvano2024,Littek2026}. A key feature of the DP framework is that it provides access to observables of all orders through a diagrammatic perturbation theory, and its structure is sufficiently flexible that substantial modifications to the underlying model can be incorporated systematically at the level of the action. 

In this work, we develop a Doi-Peliti field theory for a run-and-tumble particle in two dimensions with a completely general, static tumble distribution $\Pi(\psi)$. The distribution enters the field theory through its Fourier modes $\Pi_n$, and the resulting diagrammatic framework allows to calculate any observable exactly in terms of these modes. Interestingly, we show that all observables depending only on the $n = \pm 1$ Fourier modes, such as the MSD, mean displacements, orientation autocorrelators and position-orientation cross-correlators, can be mapped onto those of a chiral active Brownian particle (cABP) \cite{Sevilla2016}, with an effective rotational diffusivity $D_{\theta} = \alpha (1-\Re{\Pi_1})$ and an effective chirality $\Omega = - \alpha \Im{\Pi_1}$ determined by the symmetric and antisymmetric parts of the distribution respectively. Higher-order spatial moments probe higher Fourier modes, providing a systematic and complete characterisation of the tumble statistics from translational dynamical measurements alone. 

This work specifically builds on and extends the DP field theory for RTPs developed in \cite{Zhang2022a,Garcia-Millan2021}, and complements a recent study of non-uniform tumble distributions \cite{Olsen2024}, in which authors computed the MSD and orientation autocorrelator using a transfer matrix approach. The present framework yields further observables of interest, including the full hierarchy of even spatial moments, and an explicit connection to chiral ABP dynamics.  In addition to the framework making all observables of interest accessible, the flexible nature of the field-theory means that our framework can be naturally extended to interacting many-body systems \cite{Pruessner2025}, providing a foundation for future work on collective phenomena in RTPs with non-trivial tumble statistics. 

The paper is organized as follows. In Section \Sref{model}, we define the model and derive the Doi-Peliti field theory for a two-dimensional RTP with general tumble distribution, identifying the propagator $\mathcal{T}_n(\kvec,\omega)$ as the central object encoding the tumble distribution. In Section \Sref{obs}, we calculate an array of single-particle observables: the mean squared displacement (Sec.\,\ref{sec:msd}), mean displacements (Sec.\,\ref{sec:mean_disp}), orientation autocorrelators (Sec.\,\ref{sec:orientautocorr}), position-orientation cross-correlators (Sec.\,\ref{sec:posorientcrosscorr}), and higher-order spatial moments (Sec.\,\ref{sec:x4}). Section \Sref{analysis} illustrates the framework using wrapped Gaussian and bimodal Gaussian tumble distributions, demonstrating explicit control over effective persistence and chirality. We conclude in Section \Sref{conc} with a discussion of results and outlook for future work. Finally, Appendix \ref{sec:cabps} derives the field theory for chiral ABPs and makes the equivalence with the RTP model explicit at the level of the propagators. Appendix \ref{sec:DD} extends the framework to $d$ dimensions and derives the MSD for the von Mises-Fisher tumble distribution.

\section{Model and Field Theory}
\label{sec:model}

In the following, we consider an RTP moving in two dimensions, described by its position vector $\xvec(t)\in \mathbb{R}^2$ and orientation $\theta(t)\in [-\pi, \pi]$ at time $t\in \mathbb{R}$. The position of the particle evolves according to the following Langevin equation 
\begin{align}
    \dot{\mathbf{x}}(t) = \mathbf{v}_{\theta}(t) + \sqrt{2D}\boldsymbol{\eta}(t)
\end{align}
with $\vvec_{\theta}$ the velocity vector of constant magnitude $v$ that acts along the director $\dvec$
\begin{equation}
    \vvec_{\theta}(t) = v\dvec(\theta(t)),
\end{equation}
where we have defined
\begin{equation}
    \dvec(\theta(t)) = \binom{\cos(\theta(t))}{\sin(\theta(t))},
\end{equation}
and translational diffusion characterized by diffusivity $D$ and a Gaussian white noise $\bm{\eta}$
\begin{subequations}
\begin{gather}
    \langle \bm{\eta}(t) \rangle = \mathbf{0},\\
    \langle \bm{\eta}(t)\boldsymbol{\eta}^{\transpose}(t') \rangle = \mathbbm{1}_2 \delta(t-t').
\end{gather}
\end{subequations}
with $\mathbbm{1}_2$ the $2 \times 2$ identity matrix. The orientation experiences tumbles with Poissonian rate $\alpha$, and upon tumbling the particle's orientation transitions from $\theta'$ to $\theta$ with probability $\Pi(\theta' \rightarrow \theta) = \Pi(\psi = \theta-\theta')$.   
Without any loss of generality, we will consider that the particle is initialized in $\xvec_0=\mathbf{0}$ with initial orientation $\theta_0$.

\subsection{From the Fokker-Planck equation to the action}
\label{sec:FP2action}

Previous studies have explored field-theoretic approaches to deriving the dynamics and statistics of RTPs with uniform tumble statistics \cite{Garcia-Millan2021,Zhang2022a,Roberts2022}, leading to $\Pi(\psi)=1/2\pi$, and more recently in the case of a diffusive tumble angle \cite{Britton2025}. Here, the Fokker-Planck equation reads
\begin{equation}
    \partial_t P({\bf{x}}, \theta, t) = \Dx \nabla^{2}_{\bf{x}}P({\bf{x}}, \theta, t) - v\dvec(\theta)\cdot\nabla_{\bf{x}} P({\bf{x}}, \theta, t) - \alpha P({\bf{x}}, \theta, t) + \alpha \int_{-\pi}^{\pi} \plaind\psi\, \Pi(\psi)P({\bf{x}}, \theta-\psi, t),
\end{equation}
where $P({\bf x}, \theta, t)$ is the probability density of finding a particle at position $\bf{x}$ with orientation $\theta$ at time $t$. The terms, read from left to right, come from the translational diffusion with diffusivity $\Dx $, self propulsion with speed $v$ along the vector $\dvec$ and loss and gain terms from tumbling respectively away from and to the orientation prescribed by $\theta$. The action $\mathcal{A}$ is derived via the standard coherent-state path integral representation of the master equation with the Doi shift \cite{Pruessner2025}; it is classically split into a bilinear part $\mathcal{A}_{0}$ and perturbative part $\mathcal{A}_{\textrm{pert}}$ such that $\mathcal{A} = \mathcal{A}_0 + \mathcal{A}_{\textrm{pert}}$. Denoting  $\chi$ is the annihilation field and $\tilde{\chi} = 1 + \chi^{\dagger}$ is the Doi-shifted creation field \cite{Doi1976b,Cardy2008}, the bilinear and perturbative parts of the action take the following form
\begin{subequations}
\begin{align}
    \mathcal{A}_0 &= \int \plaind t \,\plaind^{2}\xvec \int_{-\pi}^{\pi} \plaind\theta\, \tilde{\chi}({\bf{x}},\theta,t)[\partial_t -\Dx \nabla_{\bf{x}}^{2} + \alpha + r]\chi({\bf{x}},\theta,t),\\
    \mathcal{A}_{\textrm{pert}} &= \int \plaind t \, \plaind^{2}\xvec \int_{-\pi}^{\pi} \plaind\theta\,\tilde{\chi}({\bf{x}},\theta,t) \biggl[v\dvec(\theta) \cdot \nabla_{\bf{x}}\chi({\bf{x}},\theta,t) - \alpha\int_{-\pi}^{\pi} \plaind \psi\, \Pi(\psi)\chi({\bf{x}}, \theta-\psi, t)\biggr],
\end{align}
\end{subequations}

All operators found in the square brackets in $\mathcal{A}_{0}$ and $\mathcal{A}_{\textrm{pert}}$ act to the right, i.e.\ act on the annihilation field $\chi$. The perturbative part of the action, $\mathcal{A}_{\textrm{pert}}$, contains only non-linear couplings, namely due to the self-propulsion $v\dvec(\theta)$ of the particle and the gain from tumbles from other states. The bilinear part of the action, $\mathcal{A}_0$, contains all other bilinear couplings. The additional mass term $r>0$ in $\mathcal{A}_0$ is a spontaneous death rate that ensures causality and the convergence of the path integral, and will be taken to be $0$ when calculating observables. The expectation of any observable $\bullet$ can be calculated via the path integral \cite{Doi1976a,Peliti1985,Cardy2008,Taeuber2014,Taeuber2005}:
\begin{equation}
    \langle \bullet \rangle = \int \mathcal{D}[\chi, \tilde{\chi}]\bullet e^{-\mathcal{A}} = \langle \bullet e^{-\mathcal{A}_{\textrm{pert}}} \rangle_{0},
\end{equation}
where $\langle \bullet \rangle_0$ is the expectation of $\bullet$ with respect to the bilinear part of the action, i.e.\
\begin{equation}
    \langle \bullet \rangle_{0} = \int \mathcal{D}[\chi, \tilde{\chi}]\bullet e^{-\mathcal{A}_{0}}.
\end{equation}
We use the following convention for the Fourier transforms of our annihilation and creation fields
\begin{subequations}
\begin{align}
    \chi({\bf{x}},\theta,t) &= \int \dbar^{2}\kvec\,\dbar\omega\,\sum_{n=-\infty}^{\infty} e^{-\imag\omega t} e^{\imag{\bf{k}}\cdot{\bf{x}}} e^{\imag n\theta}\chi_{n}({\bf{k}},\omega),\\
    \tilde{\chi}({\bf{x}},\theta,t) &= \frac{1}{2\pi}\int \dbar^{2}\kvec\,\dbar\omega\,\sum_{n=-\infty}^{\infty} e^{-\imag\omega t} e^{\imag{\bf{k}}\cdot{\bf{x}}} e^{-\imag n\theta}\tilde{\chi}_{n}({\bf{k}},\omega),
\end{align}
\label{eq:FTs_convention}
\end{subequations}
where $\mathbf{k} = [k_x,k_y]^{\transpose}$, $\dbar \omega = \plaind \omega/2\pi$ and $\dbar^2 \kvec = \plaind^2\kvec/(2\pi)^2$. The conventions used in Eq.\,\eqref{eq:FTs_convention} are a compromise of convenience \cite{Britton2025} as $\tilde{\chi}$ draws on the inverse transform of $\theta$. We introduce a similar convention for the Dirac delta function, $\deltabar(\omega) = 2\pi\delta(\omega)$ and $\deltabar^{2}(\mathbf{k}) = (2\pi)^{2}\delta^{2}(\mathbf{k})$. We also write the distribution $\Pi(\psi)$ as its Fourier expansion
\begin{equation}
    \Pi(\psi) = \sum_{n=-\infty}^{\infty} c_n e^{\imag n\psi},
\end{equation}
where the coefficients $\pi_n$ are defined as
\begin{equation}
    c_n = \frac{1}{2 \pi}\int_{-\pi}^{\pi} \plaind \phi \,\Pi(\phi) e^{-\imag n\phi}
\end{equation}
For the sake of simplicity, we introduce here the normalized factors $\Pi_n \triangleq 2\pi\, c_n$, which will be used henceforth. For a uniform distribution $\Pi(\psi) = 1/2\pi$, the distribution is already written as its Fourier series and we recover exactly the field theory of an RTP with uniform tumbling angle. As a result of Fourier transforming and further decomposing $\mathcal{A}_{\textrm{pert}} = \mathcal{A}_{v} + \mathcal{A}_{\textrm{tumble}}$, comprised of terms corresponding to self-propulsion and tumbling respectively, we arrive at the Fourier transformed action
\begin{subequations}
\begin{align}
    \mathcal{A}_0 &= \int \dbar^{2}\kvec\,\dbar\omega\,\sum_{n=-\infty}^{\infty} \tilde{\chi}_{n}(-{\bf{k}},-\omega)[-\imag\omega + \Dx  k^2 + \alpha + r]{\chi_{n}}({\bf{k}},\omega),\\
    \mathcal{A}_{v} &= -\frac{v}{2\imag}\int \dbar^{2} \kvec\,\dbar\omega\,\sum_{n=-\infty}^{\infty} \tilde{\chi}_{n}(-{\bf{k}},-\omega)\bigl[(k_x + \imag k_y)\chi_{n+1}(\mathbf{k},\omega) + (k_x - \imag k_y)\chi_{n-1}(\mathbf{k},\omega)\bigr],\\
    \mathcal{A}_{\textrm{tumble}} &= -\alpha\int \dbar^{2}\kvec\,\dbar\omega\,\sum_{n=-\infty}^{\infty} \tilde{\chi}_{n}(-{\bf{k}},-\omega)\Pi_n \chi_{n}(\mathbf{k},\omega)
\end{align}
\end{subequations}

\subsection{Bare propagator and vertices}
\label{sec:bareprop}

From $\mathcal{A}_0$, we can identify the bare propagator:
\begin{subequations}
\begin{align} \elabel{bareprop}
     \langle\chi_{n}(\mathbf{k},\omega) \tilde{\chi}_{n'}({\bf{k}}',\omega')\rangle_{0} =& \frac{\deltabar^{2}(\mathbf{k}+{\bf{k}}')\deltabar(\omega+\omega')\delta_{n,n'}}{-\imag\omega + \Dx k^{2} + \alpha + r}\\
     =& G_{n}(\mathbf{k},\omega)\deltabar^{2}(\mathbf{k}+{\bf{k}}')\deltabar(\omega+\omega')\delta_{n,n'}\\
     \triangleq& \barepropX{\mathbf{k},n,\omega}{\mathbf{k}',n',\omega'}
\end{align}
\end{subequations}
where we have introduced the Feynman diagram for a bare propagator, which is read from right to left. We too introduce perturbative vertices that will be used to construct the diagrammatics that make up our observables, identified firstly from $\mathcal{A}_{v}$
\begin{subequations}
\begin{align}
    \upvert{\mathbf{k},n,\omega}{\mathbf{k}',n-1,\omega'} &\triangleq \frac{v}{2\imag} (k_x - \imag k_y)\deltabar^{2}({\bf{k}}+{\bf{k}}')\deltabar(\omega+\omega'),\\
    \downvert{\mathbf{k},n,\omega}{\mathbf{k}',n+1,\omega'}&\triangleq \frac{v}{2\imag} (k_x + \imag k_y)\deltabar^{2}({\bf{k}}+{\bf{k}}')\deltabar(\omega+\omega'),
\end{align}
\end{subequations}
where the vertices shift the index up or down by $1$ respectively, as intuitively demonstrated by the Feynman diagrams. Finally from $\mathcal{A}_{\textrm{tumble}}$, we identify our tumble vertex
\begin{equation}
\label{pertvert2}
    \tvert{\mathbf{k},n,\omega}{\mathbf{k}',n,\omega'} \triangleq \hspace{0.5em} \alpha\Pi_n \deltabar^2(\mathbf{k}+\mathbf{k}')\deltabar(\omega+\omega') \ ,
\end{equation}

\subsection{Full Propagator}
\label{sec:fullprop}

Observables can be calculated using the full propagator,
\begin{equation}
    \langle \chi({\bf{x}},\theta,t)\tilde{\chi}({\bf{x}}_{0},\theta_{0},t_{0}) \rangle \triangleq \mathcal{G}({\bf{x}}-{\bf{x}}_{0}, \theta, \theta_{0}, t-t_{0}),
\end{equation}
which is the probability density of finding a particle at position $\bf{x}$ with director $\theta$ at time $t$ given initial conditions $\mathbf{x}(t_0) = \mathbf{x}_0, \theta(t_0) = \theta_0$ which is given by conditional probability density $P(\xvec,\theta,t|\xvec_0,\theta_0,t_0)$. 
Equivalently, in reciprocal space we have
\begin{equation}
     \langle \chi_{n}({\bf{k}},\omega)\tilde{\chi}_{m}({\bf{k}}',\omega') \rangle \triangleq \mathcal{G}_{n,m}({\bf{k}},\omega) \deltabar^{2}(\mathbf{k}+\mathbf{k}')\deltabar(\omega+\omega') 
\end{equation}
 In reciprocal space, the full propagator can straightforwardly be constructed perturbatively using the bare propagator and vertices found using a systematic approach,
\begin{equation}
    \fullprop{\mathbf{k},n,\omega}{\mathbf{k}',m,\omega'} \triangleq\, \langle \chi_{n}({\bf{k}},\omega)\tilde{\chi}_{m}({\bf{k}}',\omega') \rangle = \sum_{N=0}^{\infty} \frac{1}{N!}\Big\langle \chi_{n}({\bf{k}},\omega)\tilde{\chi}_{m}({\bf{k}}',\omega') (-\mathcal{A}_{\textrm{pert}})^{N}\Big\rangle_{0},
\end{equation}
where the hollow circle in the Feynman diagram represents a sum over all diagrams with index $m$ coming in and index $n$ leaving. While the full propagator is not needed in closed form to calculate observables, it is useful to construct the following propagators
\begin{equation}\label{gluon_diagrams}
    \barepropcurlytwo{\mathbf{k},n,\omega}{\mathbf{k}',n,\omega'} \triangleq
    \barepropshorttwo{\mathbf{k},n,\omega}{\mathbf{k}',n,\omega'}+
    \ttermone{\mathbf{k},n,\omega}{\mathbf{k}',n,\omega'} +
    \ttermtwo{\mathbf{k},n,\omega}{\mathbf{k}',n,\omega'} + \ldots
\end{equation}
which can be thought of as summing over all diagrams with the same index $n$ going in and out in the absence of self-propulsion. We can identify a geometric series and write
\begin{subequations}
\begin{align}\label{gluon1}
  \barepropcurlytwo{\mathbf{k},n,\omega}{\mathbf{k}',n,\omega'}
    \hspace{0.5em}\triangleq&\hspace{0.5em} G_n(\mathbf{k},\omega)\deltabar^{2}(\mathbf{k}+\mathbf{k}')\deltabar(\omega+\omega')\sum_{i=0}^{\infty}\left[\alpha\Pi_n G_n(\mathbf{k},\omega)\right]^{i}\\
    =&\hspace{0.5em} \frac{G_n(\mathbf{k},\omega)\deltabar^{2}(\mathbf{k}+\mathbf{k}')\deltabar(\omega+\omega')}{1-\alpha\Pi_n G_n(\mathbf{k},\omega)}
\end{align}
\end{subequations}
Using the definition of $G_n$, we finally obtain
\begin{subequations}
\begin{align}\label{gluon}
    \barepropcurlytwo{\mathbf{k},n,\omega}{\mathbf{k}',n,\omega'}
    \hspace{0.5em} \triangleq &\frac{\deltabar^{2}(\mathbf{k}+\mathbf{k}')\deltabar(\omega+\omega')}{-\imag\omega +\Dx  k^2 +\alpha(1-\Pi_n) + r}\\=&\hspace{0.5em} \mathcal{T}_n (\mathbf{k},\omega)\deltabar^{2}(\mathbf{k}+\mathbf{k}')\deltabar(\omega+\omega'),
\end{align}
\end{subequations}
where we have defined: 
\begin{equation}\elabel{TnDef}
\mathcal{T}_n (\mathbf{k},\omega) = \frac{1}{-\imag\omega +\Dx  k^2 +\alpha(1-\Pi_n) + r}
\end{equation}

\section{Positional and orientational dynamics of an RTP with preferred reorientation}
\label{sec:obs}

In the following section, we use the field theory developed in the previous section to calculate spatial and orientational moments and correlators, in order to fully characterize the effect of different tumble distributions on the dynamics of the particles. In every section, we provide exact analytical expressions and confirm our analytical results with numerics focusing as an example on the particular case of the sinusoidal tumble distribution: 
\begin{equation}
\Pi(\psi) = \frac{1}{2 \pi} (1+\beta\sin(a\psi))
\label{eq:Pi_model}
\end{equation}
with $a\in\mathbb{R}$ and $0<\beta<1$ and $\psi$ wrapped on $[-\pi,\pi]$.

\subsection{Mean Squared Displacement}
\label{sec:msd}

The mean squared displacement (MSD) is defined in real space as follows
\begin{align}\elabel{eq:MSDdef}
   \langle {|\bf{x}}|^2(t) \rangle &=  \frac{1}{2\pi}\int \plaind^{2} \xvec \int_{-\pi}^{\pi} \plaind \theta  \plaind \theta_{0}\,|{\bf{x}}|^{2}\langle \chi({\bf{x}},\theta,t)\tilde{\chi}({\bf{0}},\theta_{0},0) \rangle,
\end{align}
where we exploit rotational symmetry and average a uniformly distributed initial orientation $\theta_0$. Similarly, we make use of both spatial and temporal invariance to fix both $\mathbf{x_0}=\mathbf{0}$ and $t_0 = 0$. We use the Fourier transform of the full propagator to more succinctly calculate the observable \cite{Zhang2022a}
\begin{align}
    \langle {|\bf{x}|}^2 (t) \rangle =-\int \dbar \omega\, e^{-\imag\omega t} {\nabla_{\bf{k}}^{2}}\bigg|_{{\bf{k}}={\bf{0}}}\langle\chi_{0}(\mathbf{k},\omega) \tilde{\chi}_{0}({\bf{k}}',\omega')\rangle.
\end{align}
We now consider all diagrams in $\langle\chi_{0}(\mathbf{k},\omega) \tilde{\chi}_{0}({\bf{k}}',\omega')\rangle$ with a non-zero contribution to the MSD, that is diagrams that have an index of 0 entering and leaving, that are not \textit{killed} upon taking the Laplacian in $\mathbf{k}$ at $\mathbf{k} = \mathbf{0}$. Namely, the Laplacian restricts us to diagrams with either two self-propulsion vertices or none.

\begin{figure}[t!]
    \centering
    \includegraphics[width=0.9\linewidth]{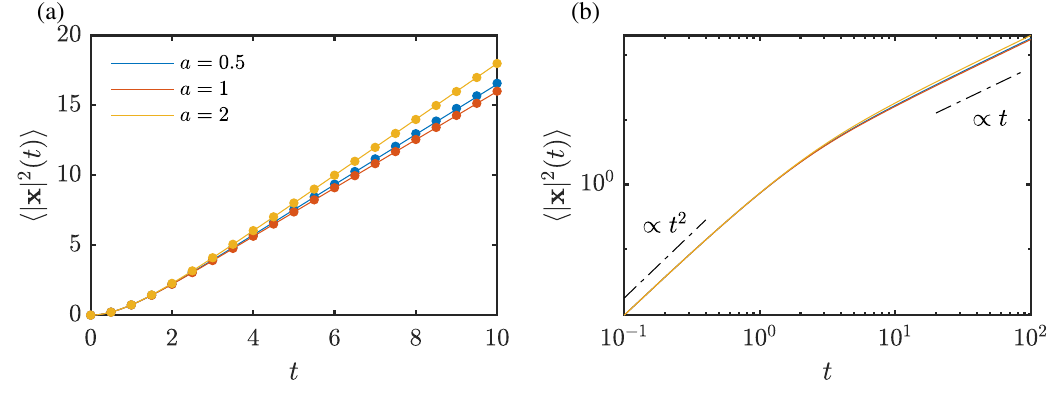}
    \caption{\textit{Mean squared displacement} -- The MSD, \Eref{MSD_2d}, of an RTP with reorientation distribution given by \Eref{Pi_model} with $a\in\{0.5,1,2\}$ and $\beta =0.8$. Here, we use the following parameters: $D=0$; $v=1$; $\alpha=1$; $\xvec_0=\nullvec$; $\theta_0=0$. The MSD is respectively plotted on both (a) a linear scale for comparison with numerical data and (b) a log-linear scale to show the distinct scaling with $t$. The symbols indicate simulation data and solid lines indicate the analytic result, \Eref{MSD_2d}. The dashed lines in panel (b) indicate respectively a ballistic behavior $\propto t^2$ and a diffusive behavior $\propto t$.} 
    \label{fig:msd}
\end{figure}

To impose an index of $n=0$ both upon entering and leaving a two-vertices diagram, the two vertices must be of opposite kind; a vertex that raises the index by 1 must be met with a vertex that lowers it by 1 and vice versa. The second-order derivatives are more trivially taken for these diagrams due to the factor of $k^2$ in the numerator, which must be hit by the Laplacian to produce non-zero contributions. We find that we can write the mean squared displacement using a vertex representation as
\begin{subequations}
\begin{align}\label{eq:msddiagrams}
    \langle |{\bf{x}}|^2 (t) \rangle &=-\int \dbar \omega\, e^{-\imag\omega t} \nabla_{\bf{k}}^{2}\biggl[\barepropcurlyshort{0}{0} + \downupprop{0}{0} + \updownprop{0}{0}\biggr]_{{\bf{k}}={\bf{0}}}\\
    &= \int \dbar \omega\, e^{-\imag\omega t} \bigl[4\Dx  \mathcal{T}_{0}({\bf 0},\omega)^{2} + v^{2}\left(\mathcal{T}_{0}({\bf 0},\omega)^{2} \mathcal{T}_{1}({\bf 0},\omega) + \mathcal{T}_{0}({\bf 0},\omega)^{2} \mathcal{T}_{-1}({\bf 0},\omega) \right) \bigr].
\end{align}
\end{subequations}
Note that here and throughout, we drop all explicit $\mathbf{k}$ and $\omega$ dependence in our diagrams when performing calculations for simplicity. All $\mathbf{k}$ and $\omega$ can be assumed to be appropriately conserved and integrated over. This implies that the MSD is only dependent on the $\theta-$modes $n=-1,0,1$. Carrying out the integral we are left with
\begin{multline}
    \langle |{\bf{x}}|^2 (t)\rangle = 4\Dx t + \frac{v^{2}}{\alpha^{2}(1-\Pi_1)^{2}}\bigl[ e^{-\alpha(1-\Pi_1) t} -1 + \alpha(1-\Pi_1)t \bigr]\\ + \frac{v^{2}}{\alpha^2(1-\Pi_{-1})^{2}}\bigl[ e^{-\alpha(1-\Pi_{-1}) t} -1 + \alpha(1-\Pi_{-1})t \bigr].
\end{multline}
Noting that $\Pi_1 = \Pi_{-1}^*$, we can rewrite the MSD in terms of the real and imaginary parts of $\Pi_1$, $\Re{\Pi_1}$ and $\Im{\Pi_1}$ respectively, as 
\begin{multline}
\elabel{MSD_2d}
    \langle |{\bf{x}}|^{2} (t) \rangle = 4\biggl[\Dx  + \frac{\alpha(1-\Re{\Pi_1})v^{2}}{2(\alpha^2(1-\Re{\Pi_1})^{2} + \alpha^2\Im{\Pi_1}^{2})} \biggr]t\\ + \frac{2v^2}{(\alpha^2(1-\Re{\Pi_1})^{2} + \alpha^2\Im{\Pi_1}^{2})^{2}}\biggl[(\alpha\Im{\Pi_1}^{2} - \alpha^2(1-\Re{\Pi_1})^{2})\bigl(1 - e^{-\alpha(1-\Re{\Pi_1})t}\cos(\alpha\Im{\Pi_1} t)\bigr)\\ - 2\alpha(1-\Re{\Pi_1}) e^{-\alpha(1-\Re{\Pi_1}) t}\alpha\Im{\Pi_1} \sin(\alpha\Im{\Pi_1} t)\biggr].
\end{multline}
Refer to \fref{msd} for an example realization. 

Interestingly, this is exactly the MSD for a chiral ABP with: 
\begin{enumerate}
\item {\it effective diffusivity} $D_\theta = \alpha(1-\Re{\Pi_1})$, controlled by the symmetric part of the tumbling distribution;
\item {\it effective chirality} $\Omega = -\alpha\Im{\Pi_1}$, controlled by the antisymmetric part of the tumbling distribution. 
\end{enumerate}
Note that the relationship between cABPs and our RTP with preferred reorientation can already be seen at the level of the bare propagator which for the RTP with preferred reorientation is given by
\begin{equation}
    \mathcal{T}_1(\mathbf{k},\omega) = \frac{1}{-\imag (\omega + \alpha\Im{\Pi_1}) + \Dx  k^2 + \alpha(1-\Re{\Pi_1}) + r} = \frac{1}{-\imag (\omega - \Omega) + \Dx  k^2 + D_\theta + r}.
\end{equation}
We thus clearly see that the bare propagator (here, with $\theta-$mode $n=1$ for instance) can generically be mapped onto that of a cABP with the above choice of rotational diffusivity $D_\theta$ and chirality $\Omega$  (see Eq.\,\eqref{eq:bareprop_cabp} in Appendix \ref{sec:cabps}). As expected, we find that the RTP with preferred reorientation displays a diffusive behavior at long times ($t \gg \alpha$) with effective diffusivity given by
\begin{align}\label{eq:deff}
    D_{\text{eff}} = \lim_{t\rightarrow \infty} \frac{\langle |{\bf{x}}|^2 (t)\rangle}{4t} = \Dx  + \frac{\alpha(1-\Re{\Pi_1})v^{2}}{2(\alpha^2(1-\Re{\Pi_1})^{2} + \alpha^2\Im{\Pi_1}^{2})}
\end{align}

\subsection{Mean Displacements}
\label{sec:mean_disp}

The equations for the mean displacement along the $x$ and $y$ axes in real space are written as follows
\begin{subequations}
\begin{align}
    \langle x(t) \rangle_{\theta_0} &= \int \plaind^{2} \xvec \int_{-\pi}^{\pi} \plaind \theta\,x\langle \chi({\bf{x}},\theta,t)\tilde{\chi}({\bf{0}},\theta_{0},0) \rangle,\\
    \langle y(t) \rangle_{\theta_0} &= \int \plaind^{2} \xvec \int_{-\pi}^{\pi} \plaind \theta\,y\langle \chi({\bf{x}},\theta,t)\tilde{\chi}({\bf{0}},\theta_{0},0) \rangle,
\end{align}
\end{subequations}
where we do not average over the initial orientation of the particle anymore as doing so would trivially lead to zero mean-displacements which we denote by $\langle \bullet \rangle_{\theta_0}$. Fourier transforming, we arrive at
\begin{subequations}
\begin{align}
    \langle x(t) \rangle_{\theta_0} &= \imag \int \dbar \omega\, e^{-\imag\omega t} \sum_{m = -\infty}^{\infty} e^{-\imag m\theta_{0}} {\partial_{k_x}\bigg|_{{\bf{k}}={\bf{0}}}\langle\chi_{0}(\mathbf{k},\omega) \tilde{\chi}_{m}({\bf{k}}',\omega')\rangle},\\
    \langle y(t) \rangle_{\theta_0} &= \imag \int \dbar \omega\, e^{-\imag\omega t} \sum_{m = -\infty}^{\infty} e^{-\imag m\theta_{0}} {\partial_{k_y}\bigg|_{{\bf{k}}={\bf{0}}}\langle\chi_{0}(\mathbf{k},\omega) \tilde{\chi}_{m}({\bf{k}}',\omega')\rangle}.
\end{align}
\end{subequations}
The partial derivatives evaluated at ${\bf{k}} = {\bf{0}}$ in the observables restrict us to diagrams containing a single self-propulsion vertex; any higher-order diagrams in $\mathbf{k}$ will be left with a prefactor of $\mathcal{O}(k_x)$ (resp. $\mathcal{O}(k_y)$) and will hence be suppressed upon evaluation at $\bm{k}=\bm{0}$. This restriction in turn allows us to identify a finite set of values of $m$ that have a non-zero contribution to the observable, despite an initially intimidating looking infinite sum. The single vertex will only ever cause an increase or decrease in the index by $1$. In turn, this means that non-zero contributions necessarily have $m = \pm 1$, in order to be brought back to $m=0$ by a single vertex. We find
\begin{subequations}
\begin{align}
    \langle x(t) \rangle_{\theta_0} &= \imag \int \dbar \omega\, e^{-\imag\omega t} \partial_{k_x}\biggl[e^{-\imag \theta_{0}} \downprop{1}{0} + e^{\imag \theta_{0}}\upprop{-1}{0}\biggr]_{{\bf{k}}={\bf{0}}} \,\\
    &= \frac{v}{2} \int \dbar \omega e^{-\imag\omega t} \bigl[e^{-\imag \theta_{0}} \mathcal{T}_{1}({\bf 0},\omega) \mathcal{T}_{0}({\bf 0},\omega) + e^{\imag \theta_{0}} \mathcal{T}_{-1}({\bf 0},\omega)\mathcal{T}_{0}({\bf 0},\omega) \bigr],\\
    \langle y(t) \rangle_{\theta_0} &= \imag \int \dbar \omega\, e^{-\imag\omega t} \partial_{k_y}\biggl[e^{-\imag \theta_{0}} \downprop{1}{0} + e^{\imag \theta_{0}}\upprop{-1}{0}\biggr]_{{\bf{k}}={\bf{0}}} \,\\
    &= \frac{v}{2\imag} \int \dbar \omega e^{-\imag\omega t} \bigl[-e^{-\imag \theta_{0}} \mathcal{T}_{1}({\bf 0},\omega) \mathcal{T}_{0}({\bf 0},\omega) + e^{\imag \theta_{0}} \mathcal{T}_{-1}({\bf 0},\omega)\mathcal{T}_{0}({\bf 0},\omega) \bigr].
\end{align}
\end{subequations}

\begin{figure}[t!]
    \centering
    \includegraphics[width=0.9\linewidth]{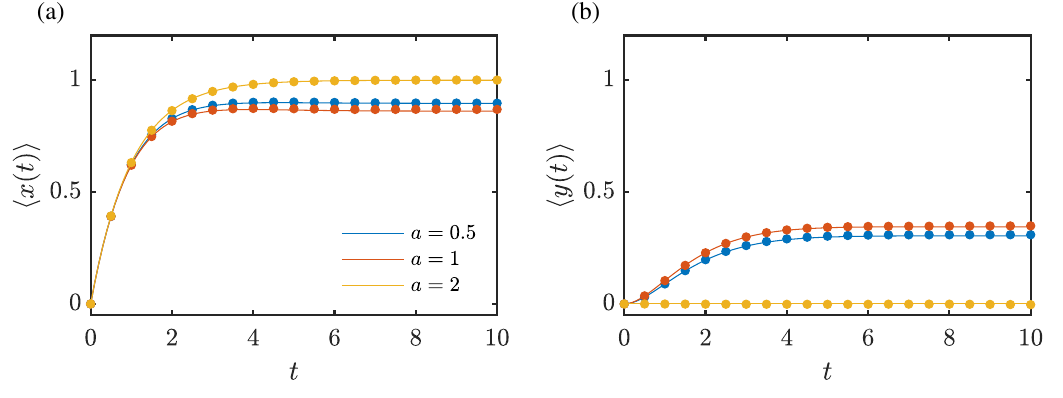}
        \caption{\textit{Mean Displacements} -- The mean displacements, \Erefs{mdx} and \eref{mdy}, of an RTP with reorientation distribution given by \Eref{Pi_model} with $a\in\{0.5,1,2\}$ and $\beta =0.8$. Here, we use the following parameters: $D=0$; $v=1$; $\alpha=1$; $\xvec_0=\nullvec$; $\theta_0=0$. Each panel respectively corresponds to (a) the mean displacement in the $x$ direction $\langle x(t)\rangle$, \Eref{mdx}, and (b) the mean displacement in the $y$ direction $\langle y(t)\rangle$, \Eref{mdy}. The symbols indicate simulation data and solid lines indicate analytic results, \Erefs{mdx} and \eref{mdy}.}
    \label{fig:md}
\end{figure}

Carrying out explicitly the integrals leaves us with
\begin{subequations}
\begin{multline}\elabel{mdx}
    \langle x(t) \rangle_{\theta_0} = \frac{v}{\alpha^2(1-\Re{\Pi_1})^2 + \alpha^2\Im{\Pi_1}^2}\bigl[-\alpha(1-\Re{\Pi_1}) e^{-\alpha(1-\Re{\Pi_1}) t} \cos(\alpha\Im{\Pi_1} t - \theta_{0})\\ + \alpha(1-\Re{\Pi_1}) \cos\theta_0 +\alpha\Im{\Pi_1} \sin\theta_0 +\alpha\Im{\Pi_1} e^{-\alpha(1-\Re{\Pi_1}) t} \sin(\alpha\Im{\Pi_1} t - \theta_{0})\bigr],
\end{multline}
\begin{multline}\elabel{mdy}
    \langle y(t) \rangle_{\theta_0} = \frac{v}{\alpha^2(1-\Re{\Pi_1})^2 + \alpha^2\Im{\Pi_1}^2}\bigl[\alpha(1-\Re{\Pi_1}) e^{-\alpha(1-\Re{\Pi_1}) t} \sin(\alpha\Im{\Pi_1} t - \theta_{0})\\ + \alpha(1-\Re{\Pi_1}) \sin\theta_0-\alpha\Im{\Pi_1} \cos\theta_0 +\alpha\Im{\Pi_1} e^{-\alpha(1-\Re{\Pi_1}) t} \cos(\alpha\Im{\Pi_1} t - \theta_{0})\bigr].
\end{multline}
\end{subequations}
Again, we numerically confirmed these analytical results as shown in \fref{md}.

\subsection{Orientation Autocorrelator}
\label{sec:orientautocorr}

We now consider the time-evolution of the director. We first calculate the orientation autocorrelator $\langle \dvec \cdot \dvec_0 \rangle$, where $\dvec_0$ is the initial director and is defined as
\begin{align}
    \dvec_0 = \binom{\cos \theta_0}{\sin \theta_0}.
\end{align}
We define the following averages $\langle \cos \theta \rangle$ and $\langle \sin \theta \rangle$ by
\begin{subequations}
\begin{align}
    \langle \cos \theta \rangle &= \int \plaind^{2}\xvec \int_{-\pi}^{\pi} \plaind \theta \,\cos \theta \langle\chi({\bf{x}},\theta,t) \tilde{\chi}({\bf{0}},\theta_{0},0)\rangle,\\
    \langle \sin \theta \rangle &= \int  \plaind^{2}\xvec \int_{-\pi}^{\pi} \plaind \theta \,\sin \theta \langle\chi({\bf{x}},\theta,t) \tilde{\chi}({\bf{0}},\theta_{0},0)\rangle.
\end{align}
\end{subequations}
By using the Fourier representation of the full propagator and exploiting the exponential form of cosine and sine, one arrives at
\begin{subequations}
\label{eq:avcosin}
\begin{align}
    \langle \cos \theta \rangle &= \frac{1}{2}\int \dbar \omega e^{-\imag\omega t}\biggl[e^{\imag \theta_{0}}\barepropcurlyshort{1}{1}+e^{-\imag \theta_{0}}\barepropcurlyshort{-1}{-1} \biggr]_{\bf{k}={\bf{0}}}\\
    &= \frac{1}{2}\int \dbar \omega e^{-\imag\omega t}\bigl[e^{\imag \theta_{0}}\mathcal{T}_{1}({\bf 0},\omega) + e^{-\imag \theta_{0}}\mathcal{T}_{-1}({\bf 0},\omega) \bigr]\\
    \langle \sin \theta \rangle &= \frac{1}{2i}\int \dbar \omega e^{-\imag\omega t}\biggl[e^{\imag \theta_{0}}\barepropcurlyshort{1}{1}-e^{-\imag \theta_{0}}\barepropcurlyshort{-1}{-1} \biggr]_{\bf{k}={\bf{0}}}\\
    &= \frac{1}{2i}\int \dbar \omega e^{-\imag\omega t}\bigl[e^{\imag \theta_{0}}\mathcal{T}_{1}({\bf 0},\omega) - e^{-\imag \theta_{0}}\mathcal{T}_{-1}({\bf 0},\omega) \bigr]
\end{align}
\end{subequations}
where the absence of higher-order terms is due to the restriction imposed by setting ${\bf{k}} = {\bf{0}}$ with no derivatives in ${\bf{k}}$, a consequence of the fact that we trivially marginalize over $\mathbf{x}$. The evolution of the director is independent of both the translational diffusion $\Dx $ and the magnitude of the velocity $v$. Computing explicitly the integrals in Eq.\,(\ref{eq:avcosin}), we conclude that 
\begin{subequations}
\begin{align}
    \langle \cos \theta \rangle &= e^{-\alpha(1-\Re{\Pi_1}) t} \cos(\theta_0 - \alpha\Im{\Pi_1} t),\\ 
    \langle \sin \theta \rangle &= e^{-\alpha(1-\Re{\Pi_1}) t} \sin(\theta_0 - \alpha\Im{\Pi_1} t). 
\end{align}
\end{subequations}
Note that the signs obtained here arise from the sign convention we took for $\Pi_1$ and the definition of $\Omega$. We can now obtain the orientation autocorrelator $\langle \dvec \cdot\dvec_0 \rangle$ in one of two ways. Our first approach uses the results already derived by rewriting the orientation autocorrelator as
\begin{align}
    \langle \dvec \cdot \dvec_0 \rangle = \cos\theta_0\langle \cos \theta \rangle + \sin\theta_0\langle \sin \theta \rangle.
\end{align}
Using simple trigonometric identities, we finally arrive at
\begin{align}
\label{eq:orientation_autocorrelator}
    \langle \dvec \cdot \dvec_0 \rangle = e^{-\alpha(1-\Re{\Pi_1}) t} \cos(\alpha\Im{\Pi_1} t).
\end{align}
Alternatively we can use a diagrammatic approach as a sanity check of our field-theoretic framework. In this case, the correlator of interest can be written using
\begin{subequations}
\begin{align}
    \langle \cos \theta\cos\theta_0 \rangle &= \frac{1}{4}\int \dbar \omega e^{-\imag\omega t}\biggl[\bigl(1 + e^{2\imag\theta_{0}}\bigr)\barepropcurlyshort{1}{1}+\bigl(1 + e^{-2\imag\theta_{0}}\bigr)\barepropcurlyshort{-1}{-1} \biggr]_{\bf{k}={\bf{0}}},\\ 
    \langle \sin \theta\sin\theta_0 \rangle &= \frac{1}{4}\int \dbar \omega e^{-\imag\omega t}\biggl[ \bigl(1-e^{2\imag\theta_{0}}\bigr)\barepropcurlyshort{1}{1}+\bigl(1 - e^{-2\imag\theta_{0}}\bigr)\barepropcurlyshort{-1}{-1} \biggr]_{\bf{k}={\bf{0}}},
\end{align}
\end{subequations}
and upon summing these, we arrive at
\begin{equation}
    \langle \dvec \cdot \dvec_0 \rangle = \frac{1}{2}\int \dbar \omega e^{-\imag\omega t}\biggl[\barepropcurlyshort{1}{1}+\barepropcurlyshort{-1}{-1} \biggr]_{\bf{k}={\bf{0}}} = e^{-\alpha(1-\Re{\Pi_1}) t}\cos(\alpha\Im{\Pi_1} t),
\end{equation}
which is consistent with Eq.\,\eqref{eq:orientation_autocorrelator}. Another interesting correlator in the context of active particles displaying chirality is given by $\langle \dvec^{\perp}\cdot\dvec_{0}\rangle$, where
\begin{align}
    \dvec^{\perp} = \binom{-\sin\theta}{ \cos\theta}
\end{align}
is the vector normal to the director $\dvec$. Similarly to what was done above, one finds in this case that
\begin{subequations}
\elabel{orientation_autocorrelator_perp}
\begin{align}
    \langle \dvec^{\perp}\cdot\dvec_{0}\rangle &= \sin \theta_0\langle \cos \theta\rangle - \cos \theta_0 \langle \sin \theta\rangle\\
    &= e^{-\alpha(1-\Re{\Pi_1}) t}\sin(\alpha\Im{\Pi_1} t),\elabel{perp_orientation}
\end{align}
\end{subequations}
As is shown in \fref{dd}, the antisymmetry, upon changing the sign of $\Im{\Pi_1}$, can be use to identify the handedness of the particle or equivalently where the antisymmetry lies in the tumbling distribution. 

\begin{figure}[t!]
    \centering
    \includegraphics[width=0.9\linewidth]{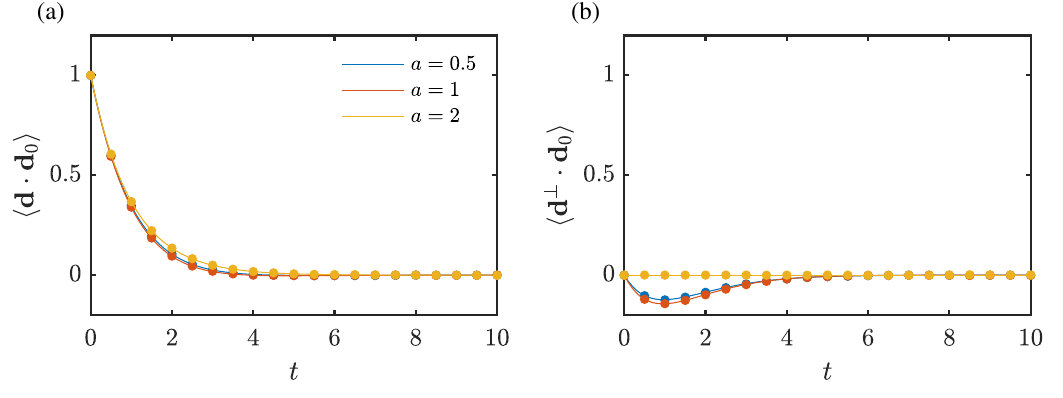}      \caption{\textit{Orientation autocorrelators} -- The orientation autocorrelators, \Erefs{orientation_autocorrelator} and \eref{perp_orientation}, of an RTP with reorientation distribution given by \Eref{Pi_model} with $a\in\{0.5,1,2\}$ and $\beta =0.8$. Here, we use the following parameters: $D=0$; $v=1$; $\alpha=1$; $\xvec_0=\nullvec$; $\theta_0=0$. Each panel respectively corresponds to (a) the orientation autocorrelator $\langle {\mathbf{d}}\cdot{\mathbf{d}}_0\rangle$,\Eref{orientation_autocorrelator} and (b) the perpendicular orientation correlator $\langle {\mathbf{d}}^{\perp}\cdot{\mathbf{d}}_0\rangle$, \Eref{perp_orientation}. The symbols indicate simulation data and solid lines indicate analytic results, \Erefs{orientation_autocorrelator} and \eref{perp_orientation}.}
    \label{fig:dd}
\end{figure}

\subsection{Position-orientation cross-correlator}
\label{sec:posorientcrosscorr}

To calculate the position-orientation cross-correlator $\langle {\bf x}\cdot\dvec\rangle$, we first consider the individual components given by 
\begin{subequations}
\begin{align}
    \langle x\cos \theta \rangle &= \frac{1}{2\pi} \int \plaind^{2} \xvec \int_{-\pi}^{\pi} \plaind \theta \plaind \theta_{0}\, x\cos \theta \langle\chi({\bf{x}},\theta,t) \tilde{\chi}({\bf{0}},\theta_{0},0)\rangle, \\
    \langle y\sin \theta \rangle &= \frac{1}{2\pi} \int \plaind^{2} \xvec \int_{-\pi}^{\pi} \plaind \theta \plaind \theta_{0}\, y\sin \theta \langle\chi({\bf{x}},\theta,t) \tilde{\chi}({\bf{0}},\theta_{0},0)\rangle. 
\end{align}
\end{subequations}
where we once again average over $\theta_0$ as in \Eref{eq:MSDdef}. Using intuition built up in calculating both positional moments and the orientation autocorrelators, we arrive at the following
\begin{subequations}
\begin{align}
    \langle x\cos \theta \rangle &= \frac{\imag}{2}\int \dbar\omega\, e^{-\imag\omega t} \partial_{k_x}\biggl[\downprop{0}{-1} + \upprop{0}{1}\biggr]_{{\bf k}={\bf 0}}\\
    &= \frac{v}{4}\int \dbar\omega\, e^{-\imag\omega t}[\mathcal{T}_{-1}({\bf 0},\omega)\mathcal{T}_{0}({\bf 0},\omega) + \mathcal{T}_{1}({\bf 0},\omega)\mathcal{T}_{0}({\bf 0},\omega)] ,\\
    \langle y\sin \theta \rangle &= -\frac{1}{2}\int \dbar\omega\, e^{-\imag\omega t} \partial_{k_y}\biggl[\upprop{0}{1} - \downprop{0}{-1}\biggr]_{{\bf k}={\bf 0}}\\
    &= \frac{v}{4}\int \dbar\omega\, e^{-\imag\omega t}[\mathcal{T}_{1}({\bf 0},\omega)\mathcal{T}_{0}({\bf 0},\omega) + \mathcal{T}_{-1}({\bf 0},\omega)\mathcal{T}_{0}({\bf 0},\omega)].
\end{align}
\end{subequations}
Subsequently, summing these results in
\begin{multline}\elabel{poss_orientation}
    \langle {\bf x}\cdot\dvec \rangle = \frac{v}{2}\int \dbar\omega\, e^{-\imag\omega t}[\mathcal{T}_{1}({\bf 0},\omega)\mathcal{T}_{0}({\bf 0},\omega) + \mathcal{T}_{-1}({\bf 0},\omega)\mathcal{T}_{0}({\bf 0},\omega)]=
    \frac{v}{\alpha^2(1-\Re{\Pi_1})^2 + \alpha^2\Im{\Pi_1}^2}\\ \times\biggl[\alpha(1-\Re{\Pi_1})+e^{-\alpha(1-\Re{\Pi_1}) t}\bigl(\alpha\Im{\Pi_1} \sin{(\alpha\Im{\Pi_1} t)} - \alpha(1-\Re{\Pi_1}) \cos{(\alpha\Im{\Pi_1} t)}\bigr)\biggr].
\end{multline}
Clearly the position-orientation cross-correlator is symmetric with respect to a change of sign in $\Im{\Pi_1}$. If we instead consider $\langle {\bf x}\cdot\dvec^{\perp} \rangle$, we can follow the same reasoning and arrive at
\begin{multline}\elabel{poss_perporr}
    \langle {\bf x}\cdot\dvec^{\perp} \rangle = \frac{\imag v}{2}\int \dbar\omega\, e^{-\imag\omega t}[\mathcal{T}_{1}({\bf 0},\omega)\mathcal{T}_{0}({\bf 0},\omega) - \mathcal{T}_{-1}({\bf 0},\omega)\mathcal{T}_{0}({\bf 0},\omega)]=
    \frac{v}{\alpha^2(1-\Re{\Pi_1})^2 + \alpha^2\Im{\Pi_1}^2}\\ \times\biggl[\alpha\Im{\Pi_1} -e^{-\alpha(1-\Re{\Pi_1}) t}\bigl(\alpha\Im{\Pi_1} \cos{(\alpha\Im{\Pi_1} t)} + \alpha(1-\Re{\Pi_1}) \sin{(\alpha\Im{\Pi_1} t)}\bigr)\biggr],
\end{multline}
which can also be used to identify the handedness of the particle. The cross-correlators are both numerically confirmed in \fref{xd}.

\begin{figure}[t!]
    \centering
    \includegraphics[width=0.9\linewidth]{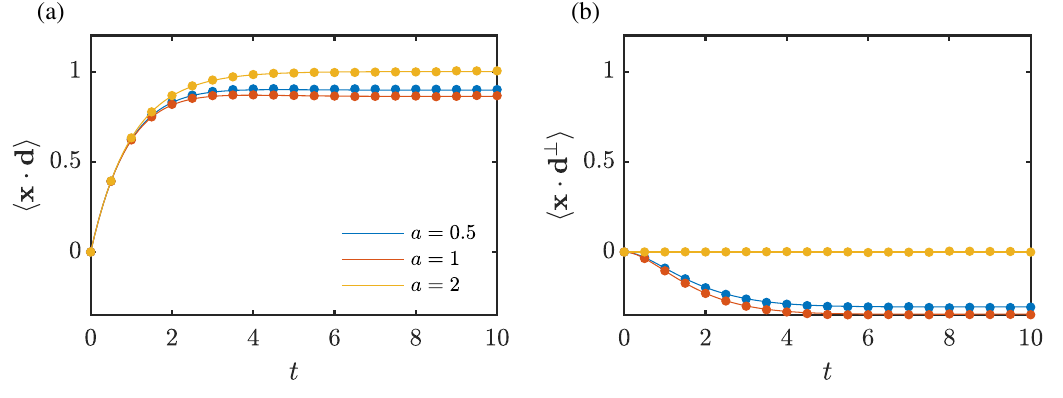}
        \caption{\textit{Position-orientation cross-correlators} -- The position-orientation cross-correlators, \Erefs{poss_orientation} and \eref{poss_perporr}, of an RTP with reorientation distribution given by \Eref{Pi_model} with $a\in\{0.5,1,2\}$ and $\beta =0.8$. Here, we use the following parameters: $D=0$; $v=1$; $\alpha=1$; $\xvec_0=\nullvec$; $\theta_0=0$. Each panel respectively corresponds to (a) the position-orientation cross-correlators $\langle \mathbf{x}\cdot{\mathbf{d}}\rangle$, \Eref{poss_orientation}, and (b) the position-perpendicular orientation cross-correlators $\langle \mathbf{x}\cdot{\mathbf{d}}^{\perp}\rangle$, \Eref{poss_perporr}. The symbols indicate simulation data and solid lines indicate analytic results, \Erefs{poss_orientation} and \eref{poss_perporr}.}
    \label{fig:xd}
\end{figure}

\subsection{\texorpdfstring{$\langle |\mathbf{x}|^4(t) \rangle$ and higher order observables}{x4(t) and higher order observables}}
\label{sec:x4}

To understand why one may be interested in going further and derive higher-order statistics for this process, we start by considering the pathological case of a Dirac comb reorientation distribution, that is a distribution comprised of $N$ Dirac-peaks uniformly spaced on $[-\pi,\pi]$, symmetric about $0$ and equally weighted
\begin{align}
    \Pi(\psi) = \frac{1}{N}\sum_{n=0}^{N-1} \delta(\psi - \psi_n) \quad \textrm{with} \quad \psi_n = \frac{2n\pi}{N} - \pi
\end{align}
For this distribution, all Fourier-modes are identically zero unless the mode is a multiple of $N$
\begin{align}
    c_m = \frac{1}{2\pi}\int_{-\pi}^{\pi}\plaind\psi\, e^{-\imag m \psi} \Pi(\psi) = \frac{1}{2\pi N}\sum_{n=0}^{N-1} e^{-\imag m \psi_n} = \left\{
\begin{array}{ll}
\frac{(-1)^m}{2\pi} & \text{if } m \equiv 0 \pmod{N} \\
    0    & \text{otherwise}
\end{array}
\right.
\end{align}

Consequently, this means that all observables calculated up until now will be entirely equivalent for particles with $N$-Dirac combs with $N\geq2$; a particle with for instance a  3-Dirac comb and a particle with 4-Dirac comb will give analytically the same lower order observables, i.e. the same MSD \Eref{MSD_2d}, motivating the pursuit of higher order observables that incorporate higher modes if one wishes to distinguish such particles.

We have already seen that for the MSD $\langle |\mathbf{x}|^{2}(t) \rangle$, only the $\pm1-$ and $0-$mode contribute to the observable, allowing us to identify the effective chirality and diffusivity of the particle. If we now consider $\langle |\mathbf{x}|^4(t) \rangle$, we can immediately see this analogy break down. In Fourier space, the equation for the fourth moment is given by
\begin{align}\label{eq:x4eq}
    \langle |\mathbf{x}|^4(t) \rangle = \int\dbar\omega e^{-\imag\omega t}\nabla_{\mathbf{k}}^{4}|_{\mathbf{k}=\mathbf{0}}\langle\chi_{0}(\mathbf{k},\omega) \tilde{\chi}_{0}({\bf{k}}',\omega')\rangle,
\end{align}
from which we can immediately identify any contributing diagram as a diagram with an even number of self-propulsion vertices less than or equal to 4, with equal numbers of up and down vertices. We find that this set of contributing diagrams is exactly the union of the set of diagrams that contributed to the calculation of the MSD, \Eref{msddiagrams}, and the set of diagrams with 4 self-propulsion vertices with incoming and outgoing $\theta-$mode $n=0$ (i.e. with 2 up-vertices and 2 down-vertices). Let $\mathfrak{D}^{(N)}$ denote the set of contributing diagrams to $\langle |\mathbf{x}|^N(t) \rangle$, and $\mathfrak{D}_N$ denote the contributing set of $N^{\textrm{th}}$ order diagrams (diagrams with $N$ vertices). One can then write
\begin{align}
    \mathfrak{D}^{(4)} = \mathfrak{D}^{(2)} \cup \mathfrak{D}_4,
\end{align}
with
\begin{align}
    \mathfrak{D}_4 \triangleq \biggl\{&\fourthpropa{\mathbf{k},0,\omega}{\mathbf{k}',0,\omega'},\fourthpropb{\mathbf{k},0,\omega}{\mathbf{k}',0,\omega'},\fourthpropc{\mathbf{k},0,\omega}{\mathbf{k}',0,\omega'},\nonumber\\&\fourthpropd{\mathbf{k},0,\omega}{\mathbf{k}',0,\omega'},\fourthprope{\mathbf{k},0,\omega}{\mathbf{k}',0,\omega'},\fourthpropf{\mathbf{k},0,\omega}{\mathbf{k}',0,\omega'}  \biggr\}.
\end{align}  

We see now that we necessarily construct diagrams that now enter the $\pm2-$modes. For a cABP, the angular velocity term in the bare propagator scales linearly with $n$ and the rotational diffusion quadratically with $n$ (see Appendix \ref{sec:cabps}). We however do not have any such relationship between $\Pi_{\pm1}$ and $\Pi_{\pm2}$, meaning that any notion of an effective chirality or diffusivity is not visible here. We see the expected short-time behavior
\begin{align}\elabel{x4short}
\langle{|\mathbf{x}|^4(t)}\rangle \simeq 32D^2t^2 + 16v^2 D t^3 + v^4 t^4 +\mathcal{O}(t^5),
\end{align}
and expected long-time behavior
\begin{align}\elabel{x4long}
\langle{|\mathbf{x}|^4(t)}\rangle \simeq 32D_\textrm{eff}^2t^2,
\end{align}
with $D_\textrm{eff}$ defined as in \Eref{deff}, once again in keeping with the fact that the particle follows Gaussian statistics at long times. This can in fact be entirely justified by writing down the excess kurtosis $\mathcal{K}(t)$, which measures how far from Gaussian the dynamics at play are, where $\mathcal{K}(t) = 0$ indicates Gaussian dynamics. At long-times, the particle's excess kurtosis is
\begin{align}
    \lim_{t\to\infty} \mathcal{K}(t) = \lim_{t\to\infty} \frac{\langle{|\mathbf{x}|^4(t)}\rangle}{2\langle{|\mathbf{x}|^2(t)}\rangle^2} - 1 \to \frac{32D_\textrm{eff}^2 t^2}{2(4D_\textrm{eff}t)^2} - 1 = 0.
\end{align}
As a result, despite this observable deviating from the cABP at the level of diagrammatics, the short- and long-time behavior indeed match.
The resulting Fourier result is numerically integrated and confirmed in \fref{x4}, where we compare to simulations.

\begin{figure}[t!]
    \centering
    \includegraphics[width=0.9\linewidth]{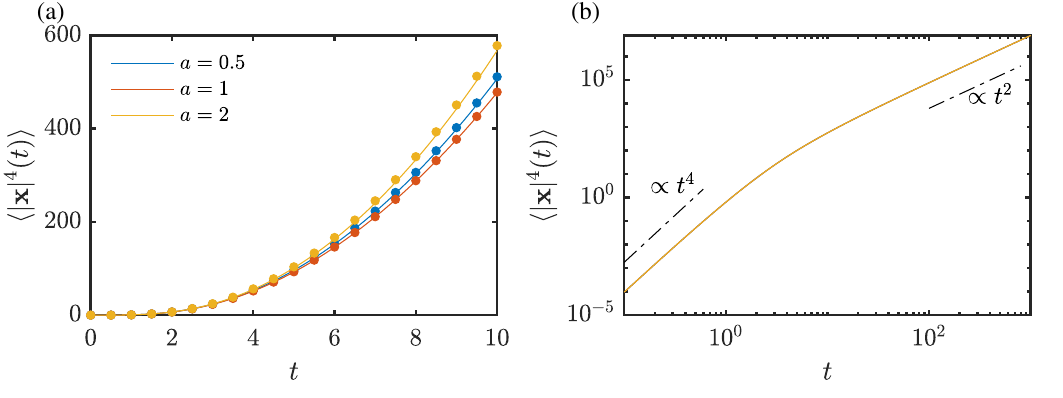}
    \caption{\textit{Fourth spatial moment} -- The fourth spatial moment, \Eref{x4eq}, of an RTP with reorientation distribution given by \Eref{Pi_model} with $a\in\{0.5,1,2\}$ and $\beta =0.8$. Here, we use the following parameters: $D=0$; $v=1$; $\alpha=1$; $\xvec_0=\nullvec$; $\theta_0=0$. The fourth moment is respectively plotted on both (a) a linear scale for comparison with numerical data and (b) a log-linear scale to show the distinct scaling with $t$. The symbols indicate simulation data and solid lines indicate the numerically integrated result from \Eref{x4eq}. The dashed lines in panel (b) indicate a transition from ballistic behavior $\propto t^4$ to diffusive behavior $\propto t^2$.} 
    \label{fig:x4}
\end{figure}

One of the key results of the framework developed in this work is that higher-order moments like $\langle |\mathbf{x}|^{2n}(t) \rangle$ are accessible for all $n$. Interestingly, we are able to learn a lot from the diagrammatics alone, including what Fourier modes contribute to observables of this form. 

In particular, calculating spatial moments of order $2n$, 
\begin{align}\elabel{x2neq}
    \langle |\mathbf{x}|^{2n}(t) \rangle = (-1)^{n}\int\dbar\omega e^{-\imag\omega t}\nabla_{\mathbf{k}}^{2n}|_{\mathbf{k}=\mathbf{0}}\langle\chi_{0}(\mathbf{k},\omega) \tilde{\chi}_{0}({\bf{k}}',\omega')\rangle,
\end{align}
involves diagrams with $2n$ self-propulsion vertices leading to all Fourier modes of index $-n$ to $n$ potentially involved in the diagrammatics. In turn, the set of all diagrams that contribute to $\langle |\mathbf{x}|^{2n}(t) \rangle$ is simply the union of the set of all diagrams that contribute to $\langle x^{2(n-1)} \rangle$ and the set of all diagrams with specifically $n$ of each self-propulsion vertex,
\begin{align}
    \mathfrak{D}^{(2n)} = \mathfrak{D}^{(2(n-1))} \cup \mathfrak{D}_{2n},
\end{align}
allowing for sequential calculation of these observables. This allows us to distinguish any and all tumble distributions from one another, as the even positional moments give a systematic way of identifying populated Fourier modes. Although we can see a clear mismatch between cABP and RTPs on the level of even moments beyond $\langle |\mathbf{x}|^{2}(t) \rangle$, we will always retain matching long-time behaviors for all moments $\langle |\mathbf{x}|^{2n}(t) \rangle$ due to the persistence of an effective diffusivity. It is known that in general
\begin{align}\elabel{x2nlong}
    \lim_{t\to\infty} \langle |\mathbf{x}|^{2n}(t) \rangle = n!(4D_{\textrm{eff}}t)^{n},
\end{align}
which can be extracted at the level of diagramatics, amounting to counting the diagrams  that only ever visit $\pm1, 0$. These are exactly the set of diagrams that produce poles of order $n+1$ after taking $k-derivatives$, which will in turn produce $t^n$ when inverse Fourier-transforming.

\section{A particular example: mixed Gaussian distributions}
\label{sec:analysis}

In the following, we analyse the effect of different wrapped Gaussian tumble distributions (also called von Mises approximation) on the dynamics of the RTP; all results are reported in \fref{gaussian}. We start with simple wrapped unimodal Gaussian distributions of the form 
\begin{align}
    \Pi(\psi) = \frac{1}{\sigma \sqrt{2\pi}} \sum_{k=-\infty}^{+\infty}\textrm{exp}{\biggl[\frac{-(\psi-\mu+2\pi k)^2}{2\sigma^2}\biggr]}.
    \label{eq:simplegaussian}
\end{align}
which we periodically extend to $[-\pi,\pi]$ and for which we will vary the mean $\mu$ and variance $\sigma^2$. 

\begin{figure}[t!]
    \centering
    \includegraphics[width=\linewidth]{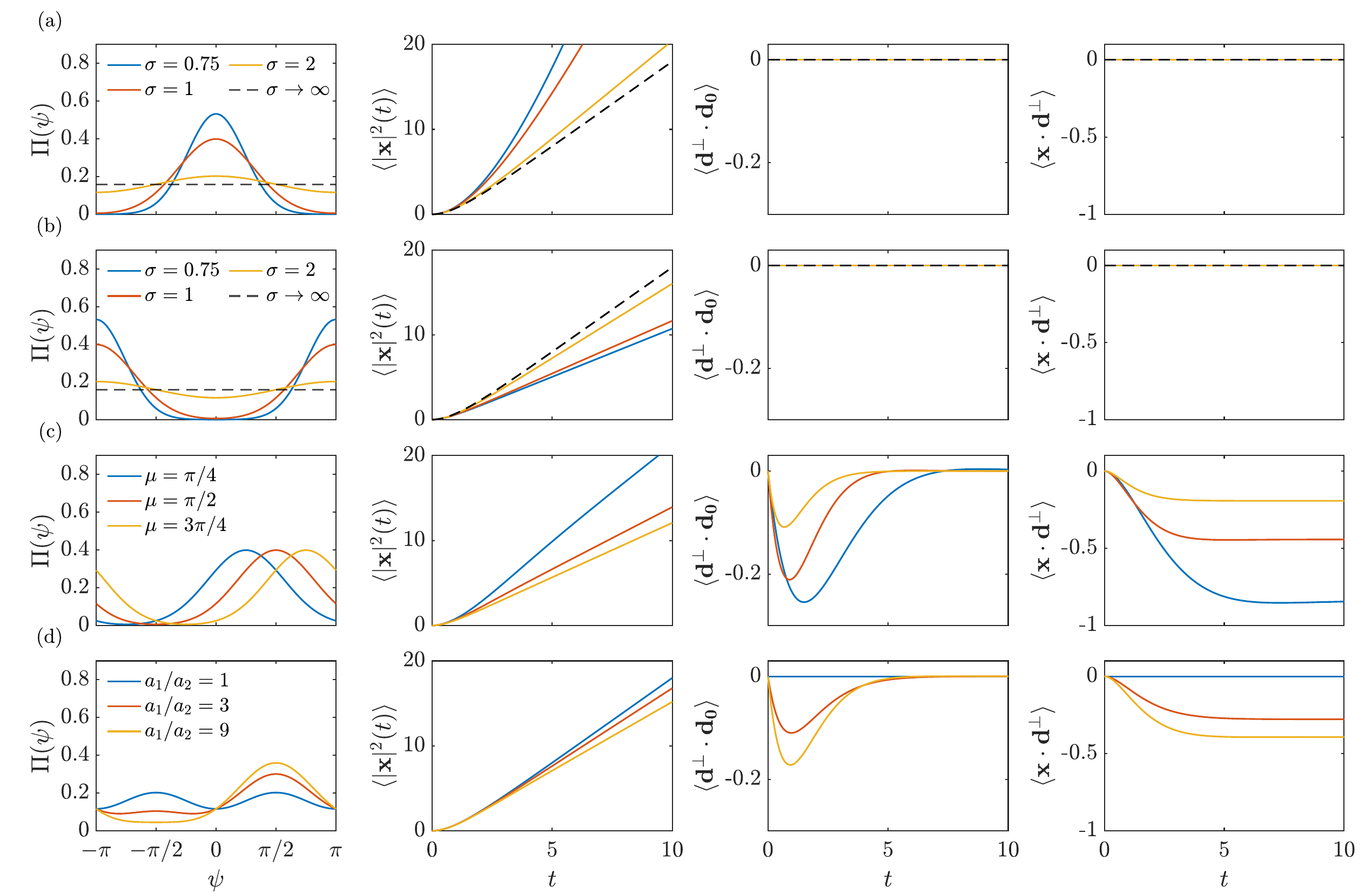}
    \caption{\textit{Gaussian Distributions} -- A variety of Gaussian tumble distributions and their corresponding observables. We plot and analyse (a) a Gaussian distribution with $\mu = 0$ and $\sigma \in \{0.75, 1, 2\}$ with comparison to $\sigma\to\infty$ (the black dashed line), which corresponds to a uniform tumble distribution; (b) a Gaussian distribution with $\mu = \pi$ and $\sigma \in \{0.75, 1, 2\}$, with comparison to $\sigma\to\infty$ (the black dashed line); (c) a Gaussian distribution with $\sigma = 1$ and $\mu \in \{\pi/4, \pi/2, 3\pi/4\}$; (d) a bimodal distribution made of unit variance Gaussian distributions centered on $\pi/2$ and $-\pi/2$ respectively with weights $a_1$ and $a_2$, the ratio of which is varied as $a_1/a_2 \in \{1,3,9\}$. For each case we plot from left to right: the distribution $\Pi(\psi)$ for $\psi\in[-\pi,\pi]$; The MSD $\langle |{\bf{x}}|^2 (t)\rangle$, \Eref{MSD_2d}; the perpendicular orientation correlator $\langle {\mathbf{d}}^{\perp}\cdot{\mathbf{d}}_0\rangle$, \Eref{perp_orientation}; the position-perpendicular orientation cross-correlator $\langle \mathbf{x}\cdot{\mathbf{d}}^{\perp}\rangle$, \Eref{poss_perporr}. For all observables, we use the further parameters: $D = 0$; $v = 1$; $\alpha = 1$; $t\in[0,10]$.}
   \label{fig:gaussian}
\end{figure}

For the distribution in \eqref{eq:simplegaussian}, the Fourier modes of index $\pm 1$ read
\begin{equation}
    \Pi_{\pm1} = e^{\mp\imag \mu - \sigma^2/2}= (\cos(\mu)\mp\imag\sin(\mu))e^{-\sigma^2/2},
\end{equation}
where again the effective persistence and chirality of the particle are respectively characterized by 
\begin{subequations}
\begin{gather}
D_{\theta} = \alpha(1-\Re{\Pi_1}) = \alpha\left(1- \cos(\mu) e^{-\sigma^2/2}\right) \\
\Omega = -\alpha\Im{\Pi_1} = \alpha \sin(\mu) e^{-\sigma^2/2}
\end{gather}
\label{eq:effectivesimplegaussian}
\end{subequations}
As seen in Fig.\,\ref{fig:gaussian}(a) and (c) and in Eq.\,\eqref{eq:effectivesimplegaussian}, we observe that increasing the variance of the distribution, both decreases the effective chirality of the particle and makes the persistence timescale converge to $\alpha^{-1}$, which is the persistence timescale of an RTP with uniform tumble distribution. The effective chirality naturally decreases as the direction of the particle is less likely to remain consistent between tumbles; said differently, a clockwise tumble is not guaranteed to be followed by a clockwise tumble as the variance increases. We note that in the case of a simple Gaussian tumble distribution centered on $\psi = 0$ or $\psi = \pi$, the observables of interest are given by 
\begin{subequations}
\begin{gather}
\langle |{\xvec}|^{2} (t) \rangle = 4\biggl[\Dx  + \frac{v^{2}}{2 \alpha (1\mp e^{-\sigma^2/2})} \biggr]t - \frac{2v^2}{\alpha^2(1\mp e^{-\sigma^2/2})^{2}}\Big[1 - e^{-\alpha(1\mp e^{-\sigma^2/2})t}\Big], \label{eq:msd_simplegaussian}\\
\langle \dvec^{\perp} \cdot \dvec_0 \rangle = 0, \\
\langle \xvec \cdot\dvec^{\perp} \rangle = 0.
\end{gather}
\end{subequations}
where the $\mu = 0$ (resp., $\mu=\pi$) case leads to minus signs (resp., plus signs) in \eqref{eq:msd_simplegaussian}. From this, we draw two main conclusions: 
\begin{enumerate}
\item increasing the variance means the tumble angle of the particle becomes less dependent on the mean, making the dynamics far more akin to RTP with a uniform tumble distribution; in the limit of a uniform distribution $\sigma \to \infty$, we recover the well-known MSD for the original RTP given by
\begin{equation}
\langle |{\xvec}|^{2} (t) \rangle = 4\biggl[\Dx  + \frac{v^{2}}{2 \alpha} \biggr]t - \frac{2v^2}{\alpha^2}\Big[1 - e^{-\alpha t}\Big];
\end{equation}
\item chirality-related observables are identically zero for distributions which are symmetric around $\psi = 0$ or $\psi=\pi$ capturing the fact that an effective chirality emerges as argued above from the asymmetry in the distribution. 
\end{enumerate}

We can confirm this by additionally looking at the effect of changing the distribution mean while keeping the variance constant (see Fig.\,\ref{fig:gaussian}(b)). First, Eq.\,\eqref{eq:effectivesimplegaussian} teaches us that changing the mean of the distribution has out-of-phase effects on the effective diffusivity (and as a result persistence) and effective chirality. Namely, both vary sinusoidally with the mean $\mu$; however, the maximum chirality is attained at $\mu =\pm \pi/2$, when the particle performs on average sharp left or right turns by $\pi/2$, while the maximum persistence is attained at $\mu = 0$ when the persistence timescale diverges. Inversely, the minimum chirality is attained at $\mu = \{-\pi, 0, \pi\}$ where no notion of left- or right-handedness can be seen on average in the tumbles, and the minimum persistence is attained at $\mu = \pm\pi$, where the particle performs on average full reversals when tumbling. When $\mu \neq \{-\pi,0,\pi\}$, the correlators  $\langle \dvec^{\perp} \cdot \dvec_0 \rangle$ and $\langle \xvec \cdot\dvec^{\perp} \rangle$ are not identically zero anymore. For instance, the former is given for simple Gaussian tumble distributions by 
\begin{equation}
\langle \dvec^{\perp} \cdot \dvec_0 \rangle = - e^{-D_{\theta}t} \sin \Omega t
\end{equation}
This correlator described the chiral memory of the particle. We thus expect this display oscillations around 0 decaying over time with a decay rate increasing with decreasing particle persistence which corresponds to increasing $\mu$ here. At long time, we expect the correlator to vanish as is observed in Fig.\,\ref{fig:gaussian}(c). Further, the correlator $\langle \xvec \cdot\dvec^{\perp} \rangle$ represents the expected transverse displacement of the particle relative to its current orientation. For a particle with non-zero effective chirality ($\mu \neq 0$), tumbles tend to curve its trajectory in one direction, leaving its historical position systematically to the left or right of its current heading. We expect it to decay monotonically with time. In the long time limit ($t \to \infty$), this reaches a steady-state constant lateral offset of
\begin{equation}
\langle \xvec \cdot\dvec^{\perp} \rangle = -\frac{v \Omega}{D_{\theta}^2 + \Omega^2} = -\frac{v}{2 \alpha} \frac{\sin \mu}{ \cosh \frac{\sigma^2}{2}-\cos \mu}
\end{equation}
whose absolute value monotonically decreases as $\mu$ moves away from $0$ as confirmed by Fig.\,\ref{fig:gaussian}(c).

We additionally consider mixed Gaussian distributions of the form
\begin{align}
        \Pi(\psi) = \frac{a_1}{\sigma_1 \sqrt{2\pi}} \sum_{k=-\infty}^{+\infty}\textrm{exp}{\biggl[\frac{-(\psi-\mu_1+2\pi k)^2}{2\sigma_1^2}\biggr]} + \frac{a_2}{\sigma_2 \sqrt{2\pi}} \sum_{k=-\infty}^{+\infty}\textrm{exp}{\biggl[\frac{-(\psi-\mu_2+2\pi k)^2}{2\sigma_2^2}\biggr]},
\end{align}
where $a_1$ and $a_2$ are the respective normalized weights of each Gaussian (such that $a_1+a_2=1$), equipped with means $\mu_{1}, \mu_{2}$ and variance $\sigma_{1}^2, \sigma_{2}^2$. Results for mixed Gaussian are reported in Fig.\,\ref{fig:gaussian}(d). In this case, the Fourier modes of index $\pm 1$ read
\begin{subequations}
\begin{align}
    \Pi_{\pm 1} &= a_1e^{\mp\imag \mu_1 - \sigma_1^2/2} + a_2e^{\mp\imag \mu_2 - \sigma_2^2/2}\\
    &= a_1(\cos(\mu_1)\mp\imag\sin(\mu_1))e^{-\sigma_1^2/2} + a_2(\cos(\mu_2)\mp\imag\sin(\mu_2))e^{-\sigma_2^2/2}.
\end{align}
\end{subequations}
Varying the ratio of their relative weights, we can increase the prevalence of one the Gaussian peaks. This subsequently results in dynamics that are governed by the variance and mean of that mode. In \fref{gaussian}, we find that for two Gaussian symmetrically positioned relative to $0$ (i.e. at $\mu_1=\pi/2, \mu_2=-\pi/2, \sigma_1^2 = \sigma_2^2$), tuning the relative weights allows us to tune the effective chirality and persistence. Increasing (equiv. decreasing) the ratio of their weights from $1$ decreases the persistence of the particle as the tumble distribution is no longer symmetric, and increases the effective chirality as tumbles have a preferential direction. 

\section{Discussion and outlook}
\label{sec:conc}

In this work, we have developed a Doi-Peliti field theory for a run-and-tumble particle in two dimensions with a completely general static tumble distribution $\Pi(\psi)$. The distribution enters the field theory exclusively through its Fourier modes ${\Pi_n}$, and the resulting diagrammatic framework provides a systematic and exact route to observables of arbitrary order.

A key result of this work is that all observables depending only on the $n = \pm 1$ Fourier modes of the tumble distribution are exactly equivalent to those of a chiral active Brownian particle (cABP). Specifically, the mean squared displacement (Eq.~\eref{MSD_2d}), mean displacements (Eqs.~\eref{mdx} and~\eref{mdy}), orientation autocorrelators (Eqs.~\eref{orientation_autocorrelator} and~\eref{orientation_autocorrelator_perp}), and position-orientation cross-correlators (Eqs.~\eref{poss_orientation} and~\eref{poss_perporr}) all depend on the tumble distribution only through the effective rotational diffusivity $D_{\theta} = \alpha(1 - \Re{\Pi_1})$ and the effective chirality $\Omega = -\alpha\Im{\Pi_1}$, controlled respectively by the symmetric and antisymmetric parts of $\Pi(\psi)$. This equivalence can be understood directly at the level of the dressed propagator $\mathcal{T}_n(\mathbf{k},\omega)$, which for $n = \pm 1$ maps exactly onto the bare propagator of a cABP (Appendix~\ref{sec:cabps}). Intuitively, the $n = 1$ Fourier mode of the tumble distribution captures the mean reorientation, which is all that is needed to determine the long-time orientational relaxation and the drift of the particle. This result has a direct practical implication: measuring any one of the observables listed above is sufficient to determine $\Pi_1$, and hence both the effective persistence timescale $D_\theta^{-1}$ and the effective chirality $\Omega$. In particular, the perpendicular orientation autocorrelator and the position-perpendicular orientation cross-correlator are sensitive to the sign of $\Im{\Pi_1}$, providing a direct experimental handle on the handedness of the tumble distribution. 

A key limitation of the cABP equivalence is that it is blind to higher Fourier modes $\Pi_n$ with $|n| \geq 2$, which only become accessible through higher-order spatial moments. We showed that the $2n$-th spatial moment $\langle |\mathbf{x}|^{2n}(t)\rangle$ (Eq.~\eref{x2neq}) depends on all Fourier modes $\Pi_m$ with $|m| \leq n$, with the set of contributing diagrams satisfying the recursive relation $\mathfrak{D}^{(2n)} = \mathfrak{D}^{(2(n-1))} \cup \mathfrak{D}{2n}$. This provides a systematic, sequential route to probing the full Fourier content of $\Pi(\psi)$: the MSD probes $\Pi{\pm 1}$, the fourth moment additionally probes $\Pi_{\pm 2}$, and so on. The fourth spatial moment (Eq.\eref{x4eq}) illustrates this clearly. While it deviates from the cABP fourth moment at intermediate times, it matches the cABP result at both short times (Eq.\eref{x4short}, where tumbles have not yet occurred) and long times (Eq.\eref{x4long}, where Gaussian diffusion is recovered with $\langle |\mathbf{x}|^4(t)\rangle \simeq 32 D_\mathrm{eff}^2 t^2$). The long-time behavior of all even moments is governed by the effective diffusivity $D_\mathrm{eff}$ alone, consistent with the central limit theorem. The approach to this Gaussian limit, however, carries information about higher Fourier modes through the transient behavior, and it is precisely this transient regime that distinguishes particles with different tumble distributions sharing the same $\Pi_{\pm 1}$. The full hierarchy of even moments therefore provides, at least in principle, a complete characterisation of the tumble distribution. Interestingly, while orientational observables are often hard to measure in experimental settings leading to inaccurate measurement of tumble distributions for instance, our field-theoretic framework allows us to reconstruct the tumble distribution one Fourier mode at a time by probing only translational moments which are experimentally accessible with very high precision. 

In the last part of our work, our framework was illustrated concretely with wrapped Gaussian tumble distributions. We showed that the effective chirality and persistence vary out of phase as functions of $\mu$; indeed, we showed that chirality is maximal at $\mu = \pm\pi/2$ while persistence is maximal at $\mu = 0$ and minimal at $\mu = \pi$. This phase relationship implies a physically relevant trade-off: a tumble distribution cannot simultaneously maximise both persistence and chirality. Increasing the variance $\sigma^2$ drives both $D_\theta \to \alpha$ and $\Omega \to 0$, recovering the dynamics of an RTP with uniform tumble distribution in the limit $\sigma \to \infty$, as the broadening distribution approaches uniformity and washes out any preferred direction. Our bimodal Gaussian example further demonstrates that the effective dynamics can be tuned continuously by varying the relative weights of two Gaussian components, providing a concrete mechanism for interpolating between chiral and achiral dynamics. These results connect directly to the biological context motivating this work. The tumble angle distribution of \textit{E. coli} measured by Berg and Brown \cite{Berg1972} peaks near $68^\circ$, corresponding to a wrapped Gaussian with $\mu \approx 1.19$ rad, producing a particle with non-trivial persistence ($D_\theta < \alpha$) and, if the distribution is asymmetric, a non-zero effective chirality. The latter could arise for instance in the presence of a chemical gradient, where tumble angles are known to be biased \cite{Saragosti2012}. The ability to directly read off $D_\theta$ and $\Omega$ from the $\pm 1$ Fourier modes of the measured distribution again provides a quantitative bridge between single-cell tracking experiments and effective particle models.

This work opens several directions for future investigation. As argued above, the hierarchy of spatial moments provides a systematic route to inferring the Fourier modes of $\Pi(\psi)$ from experimental trajectory data, with the $\pm 1$ modes accessible from the MSD and higher modes requiring higher-order moments. Developing a practical inference scheme, which accounts for finite trajectory length and measurement noise, would be a valuable contribution to the analysis of single-cell tracking data, and could provide a means of directly measuring non-uniform tumble distributions in bacteria from trajectory statistics alone. The Doi-Peliti framework developed here also provides a natural starting point for the analysis of interacting RTPs with non-uniform tumble distributions. In the uniform case, motility-induced phase separation (MIPS) is a well-established collective phenomenon \cite{Cates2015}; whether the effective chirality $\Omega$ introduced by an asymmetric tumble distribution suppresses MIPS, as is the case for chiral ABPs \cite{Ma2022,Liebchen2022}, and how this depends on the full Fourier content of $\Pi(\psi)$, remains an open question. Beyond this, extending the framework to accommodate a tumble distribution $\Pi(\psi, \mathbf{x}, t)$ that varies in space and time would be a significant step towards a field-theoretic description of chemotactic navigation, where the tumble rate and tumble angle distribution are modulated by the local chemical environment. Finally, while the $d$-dimensional framework developed in Appendix~\ref{sec:DD} provides the MSD for RTPs with isotropic tumble distributions in arbitrary dimension, a full treatment of anisotropic tumble distributions in three dimensions, relevant for the 3D motility of \textit{E. coli} as characterised experimentally by Figueroa-Morales et al. \cite{Figueroa-Morales2020}, would require extending our analysis and we leave it for future work.

\section*{Acknowledgments}
The authors would like to thank Gunnar Pruessner for interesting conversations. CB was supported by a Roth PhD scholarship funded by the Department of Mathematics at Imperial College London. ZZ was supported by Fujian Provincial Excellent Postdoctoral Program.

\appendix

\section{Field Theory of chiral active Brownian particles (cABPs)}
\label{sec:cabps}

In the following we consider an chiral ABP moving in two dimensions, described by its position vector ${\bf{x}}(t)$ and orientation $\theta(t)$ at time $t$. Following the notation of the main text, the dynamics of the particle can be described by the following Langevin equations:
\begin{subequations}
\begin{align}
    \dot{\bf{x}}(t) &= \vvec_\theta(t) + \sqrt{2\Dx}{\bm{\eta}}(t),\\
    \dot{\theta}(t) &= \Omega + \sqrt{2D_\theta}\xi(t).
\end{align}
\end{subequations}
The particle's position and orientation are subject to diffusion with diffusivities $D$ and $D_\theta$ respectively, and both ${\bm{\eta}}$ and $\xi$ are Gaussian white noise with means $\langle {\bm{\eta}}(t)\rangle = \bf{0}$, $\langle {\xi(t)}\rangle = 0$ and auto-correlation functions $\langle \bm{\eta} (t) \bm{\eta}^{\transpose} (t') \rangle = \mathbbm{1}_2 \delta(t - t')$, $\langle \xi (t) \xi (t') \rangle = \delta(t - t')$. The velocity here has constant magnitude and acts along the director $\dvec$
\begin{equation}
    {\bf{v}_{\theta}}(t) = v\dvec(\theta(t)) = v\binom{\cos{\theta}}{ \sin{\theta}}.
\end{equation}
The orientation additionally evolves with an angular velocity $\Omega$, where the particle is left- or right-handed depending on the sign of $\Omega$; $\Omega>0$ results in left-handed trajectories and $\Omega<0$ results in right-handed trajectories.
The Fokker-Planck equation for a cABP in two dimensions reads
\begin{equation}
    \partial_t P({\bf{x}}, \theta, t) = D_\theta \partial_{\theta}^{2} P({\bf{x}}, \theta, t) + \Dx\nabla^{2}_{\bf{x}}P({\bf{x}}, \theta, t) - v\dvec\cdot\nabla_{\bf{x}} P({\bf{x}}, \theta, t) - \Omega \partial_\theta P({\bf{x}}, \theta, t) ,
\end{equation}
where $P({\bf x}, \theta, t)$ is the probability density of finding a particle at position $\bf{x}$ with orientation $\theta$ at time $t$. Following the same procedure as highlighted in the main text, the action $\mathcal{A}=\mathcal{A}_{0}+\mathcal{A}_{\text{pert}
}$ follows immediately \cite{Pruessner2025} and can be Fourier-transformed to give 
\begin{subequations}
\begin{align}
    \mathcal{A}_0 &= \int \dbar^{2}\kvec\dbar\omega\,\sum_{n=-\infty}^{\infty} \tilde{\chi}_{n}(-{\bf{k}},-\omega)[-\imag\omega + D_\theta n^{2} + \Dx k^2 + \imag n\Omega + r]{\chi_{n}}({\bf{k}},\omega),\\
    \mathcal{A}_{\textrm{pert}} &= -\frac{v}{2\imag}\int \dbar^{2}\kvec\dbar\omega\,\sum_{n=-\infty}^{\infty} \tilde{\chi}_{n}(-{\bf{k}},-\omega)\bigl[(k_x + \imag k_y)\chi_{n+1}(\mathbf{k},\omega) + (k_x - \imag k_y)\chi_{n-1}(\mathbf{k},\omega)\bigr].
\end{align}
\end{subequations}
From $\mathcal{A}_0$, we can identify the bare propagator:
\begin{subequations}
\begin{align} \label{eq:bareprop_cabp}
     \langle\chi_{n}(\mathbf{k},\omega) \tilde{\chi}_{m}({\bf{k}}',\omega')\rangle_{0} =& \frac{\deltabar^{2}(\mathbf{k}+{\bf{k}}')\deltabar(\omega+\omega')\delta_{n,m}}{-\imag(\omega-n\Omega) + D_{\theta}n^{2} + \Dx k^{2} + r}\\
     =& G_{n}(\mathbf{k},\omega)\deltabar^{2}(\mathbf{k}+{\bf{k}}')\deltabar(\omega+\omega')\delta_{n,n'}\\
     \triangleq& \barepropX{\mathbf{k},n,\omega}{\mathbf{k}',n',\omega'}\ .
\end{align}
\end{subequations}
 We too find the same self-propulsion vertices in $\mathcal{A}_{\textrm{pert}}$:
\begin{subequations}
\begin{align}
    \upvert{n}{n-1} &\triangleq \frac{v}{2\imag} (k_x - \imag k_y)\deltabar^{2}({\bf{k}}+{\bf{k}}')\deltabar(\omega+\omega'),\\
    \downvert{n}{n+1} &\triangleq \frac{v}{2\imag} (k_x + \imag k_y)\deltabar^{2}({\bf{k}}+{\bf{k}}')\deltabar(\omega+\omega').
\end{align}
\end{subequations}
We can clearly see equivalence between cABPs and RTPs with preferred rotation at the level of the bare propagators, \Erefs{bareprop_cabp} and \eref{TnDef}, \textit{given} the $\theta-$mode $n\in\{-1,0,1\}$ with $D_\theta = \alpha(1-\Re{\Pi_1})$ and $\Omega = -\alpha\Im{\Pi_1}$; any observable that is exclusively dependent on these $\theta-$modes will show equivalence between the two models.

\section{\texorpdfstring{$d$-dimensional field theory and MSD}{d-dimensional field theory and MSD}}
\label{sec:DD}
In the following, we expand our framework into $d$-dimensional space. In this case, the director is hopping on a $d$-dimensional unit sphere. We introduce the transition probability as
\begin{equation}
    \Pi(\Theta'\rightarrow \Theta)=\Pi(\dvec_{\Theta'}\cdot\dvec_\Theta)\ ,
\end{equation}
assuming a non-chiral tumble distribution such that we can subsequently use a Gegenbauer transformation, where we use the inner product $\dvec_{\Theta}\cdot\dvec_{\Theta'}=\cos{\tilde{\theta}}$ to represent the angle $\tilde{\theta}$ between two unit vectors. Here, $\dvec_\Theta$  
is the unit vector indicating the direction of motion in $d$-dimensions
\begin{equation}
    \dvec_\Theta=\begin{bmatrix}
\sin{\theta_1} \sin{\theta_2} \dots\sin{\theta_{d-3}}\sin{\theta_{d-2}} \sin\psi \\
\sin{\theta_1} \sin{\theta_2} \dots\sin{\theta_{d-3}}\sin{\theta_{d-2}} \cos\psi \\
\sin{\theta_1} \sin{\theta_2} \dots\sin{\theta_{d-3}}\cos{\theta_{d-2}}  \\
\dots\\
\sin{\theta_1} \cos{\theta_2} \\
\cos{\theta_1}
\end{bmatrix} \ .
\end{equation}

The Fokker-Planck equations now reads
\begin{multline}
    \partial_t P({\bf{x}}, \Theta, t) = \Dx \nabla^{2}_{\bf{x}}P({\bf{x}}, \Theta, t) - v\dvec_{\Theta}\cdot\nabla_{\bf{x}} P({\bf{x}}, \Theta, t) - \alpha P({\bf{x}}, \Theta, t)\\+ \alpha \int d\Theta'\, \Pi(\Theta'\rightarrow\Theta)P({\bf{x}}, \Theta', t)\ .
\end{multline}
To obtain the statistics of the motion of RTPs in a $d$-dimensional space we first need to introduce the hyperspherical harmonics \cite{Wen1985}; these are the eigenfunctions of the operator $\nabla^2_\Theta$ in $d$-dimensional space such that
\begin{equation}
    -\nabla^2_\Theta Y_\ell^{m_1,m_2\dots m_{d-2}}(\Theta)=\ell(\ell+d-2)Y_\ell^{m_1,m_2\dots m_{d-2}}(\Theta)\triangleq\ell(\ell+d-2)Y_\ell^{\mvec}(\Theta) \ ,
\end{equation}
where $\mvec=[m_1,m_2\dots m_{d-2}]$. When $d=3$, the hyperspherical harmonics are exactly the well-known spherical harmonics $Y_\ell^m$. These special functions have the following orthogonality properties 
\begin{equation}
    \int \dint \Theta Y_\ell^{\mvec}(\Theta)\Tilde{Y}_{\ell'}^{\mvec'}(\Theta)=\delta_{\ell,\ell'}\delta_{\mvec,\mvec'}\ , 
\end{equation}
where $\Tilde{Y}_{\ell'}^{\mvec'}(\Theta)=(Y_{\ell'}^{\mvec'}(\Theta))^\dagger$ represents the complex conjugate, and the surface element is
\begin{equation}
    \dint\Theta=(\sin{\theta_1})^{d-2}(\sin{\theta_2})^{d-3}\ldots\sin{\theta_{d-2}}\,\dint{\theta_1}\dint{\theta_2}\dots\dint{\theta_{d-2}}\dint\psi \ ,
\end{equation}
with $\theta_j\in[0,\pi)$ and $\psi\in[0,2\pi)$. The surface area of a unit $d$-sphere is given by
\begin{equation}
S_d \triangleq   \int \dint \Theta_d =\frac{2\pi^{d/2}}{\Gamma(d/2)} \ ,
\end{equation}
where $\Gamma(x)$ is the Gamma function.

\subsection{Field theory}
Similar to the $d=2$ case, \Sref{model}, we start from with the Fokker-Planck equation and derive the action
\begin{subequations}
    \begin{align}
        \action&=\action_0+\action_v+\mathcal{A}_{\textrm{tumble}}\ ,\\
        \elabel{d_action_bilinear}
        \action_0(\chi,\tilde{\chi})&=\int \dint t\int \ddint {\xvec}\int\dint{\Theta}\tilde{\chi}(\xvec,\Theta,t)\left[\partial_t-D\nabla_\xvec^2+\tumbleRate + r\right]\chi(\xvec,\Theta,t) \ ,\\
        \elabel{d_dimensional_w_pert}
         \action_v(\chi,\tilde{\chi})&=v\int \dint t\int \ddint {\xvec}\int\dint{\Theta}\tilde{\chi}(\xvec,\Theta,t)\dvec_\Theta\cdot\nabla_\xvec\chi(\xvec,\Theta,t) \ , \\
          \elabel{d_action_tumble}\mathcal{A}_{\textrm{tumble}}(\chi,\tilde{\chi})&=-\tumbleRate\int \dint t\int \ddint {\xvec}\int\dint{\Theta}\tilde{\chi}(\xvec,\Theta,t)\int\dint\Theta'\chi(\xvec,\Theta',t)\Pi(\dvec_{\Theta}\cdot\dvec_{\Theta'}) \ ,
 \end{align}
\end{subequations}
where $\xvec=[x_1,x_2\dots x_{d-1}, x_d]^T$. We further introduce the fields as
\begin{subequations}
\elabel{d_fields}
    \begin{align}
        \chi(\xvec,\Theta,t)&=\int\dbar{\omega}\exp{-\imag \omega t}\int\ddintbar{\kvec}\exp{\imag \kvec\cdot\xvec}\sum_{\ell,\mvec} Y_\ell^\mvec(\Theta)\chi_\ell^\mvec(\kvec,\omega) \ ,\\
    \tilde{\chi}(\xvec,\Theta,t)&=\int\dbar{\omega}\exp{-\imag \omega t}\int\ddintbar{\kvec}\exp{\imag \kvec\cdot\xvec}\sum_{\ell,\mvec} \Ytilde_\ell^\mvec(\Theta)\tilde{\chi}_\ell^\mvec(\kvec,\omega) \ ,
    \end{align}
\end{subequations}
where $\kvec=[k_1,k_2,\dots k_{d-1},k_d]^T$.
Substituting \Eref{d_fields} into the bilinear action \Eref{d_action_bilinear} first, we obtain the bare propagator 
\begin{equation}
\barepropX{\kvec,\omega,\ell,\mvec}{\kvec',\omega',\ell',\mvec'} \triangleq \frac{\delta_{\ell,\ell'}\delta_{\mvec,\mvec'}\deltabar^d(\kvec+\kvec')\deltabar(\omega+\omega')}{-\imag\omega+D k^2+\alpha+r}  .
\end{equation}
Using both the Gegenbauer transformation and the addition theorem for hyperspherical harmonics \cite{Wen1985} below
\begin{equation}
    \Gegenbauer{d}_\ell(\dvec_\Theta\cdot \dvec_{\Theta'})=\frac{S_d (d-2)}{(2\ell +d -2)}\sum_{\mvec} Y_\ell^\mvec(\Theta) \Ytilde_\ell^\mvec(\Theta') \ ,
\end{equation}
the perturbative part of the action with respect to the tumble event \Eref{d_action_tumble} can be written as 
\begin{equation}
    \mathcal{A}_{\textrm{tumble}}=-\alpha \int\dbar\omega\dbar{\omega'}\int\ddintbar{\kvec}\ddintbar{\kvec'}\sum_{\ell,\mvec}\tilde{\chi}_\ell^\mvec(\kvec',\omega')\chi_\ell^\mvec(\kvec,\omega)\Pi^{(d)}_\ell\deltabar(\omega+\omega')\deltabar^d(\kvec+\kvec') \frac{S_d (d-2)}{(2\ell +d -2)} \ ,
\end{equation}
where the transform of the distribution $\Pi(x)$ takes the form
\begin{subequations}
\elabel{Gegenbauer_transform}
\begin{equation}
    \begin{gathered}
        \Pi(x)=\sum_{\ell=0}^\infty \Pi^{(d)}_\ell \Gegenbauer{d}_\ell(x) \ ,\\
        \Pi^{(d)}_\ell= \frac{\ell! (\ell+d/2-1)\Gamma^2(d/2-1)}{\pi 2^{3-d}\Gamma(\ell+d-2)} 
        \int_{-1}^1\dint{x} (1-x^2)^{(d-3)/2} \Gegenbauer{d}_\ell(x) \Pi(x) \ ,
    \end{gathered}
\end{equation}
\end{subequations}
where $\Gamma$ is the well-known Gamma function, $\Gegenbauer{d}_\ell(x)$ is the Gegenbauer polynomials. A perturbative vertex capturing the tumble events is produced as 
\begin{equation}
    \elabel{d_tumble_vertex}
     \tumblepert{\ell',\mvec'}{\ell,\mvec}\triangleq\alpha\frac{S_d (d-2)}{(2\ell +d -2)}\Pi^{(d)}_\ell \delta_{\ell,\ell'}\delta_{\mvec',\mvec} \ .
\end{equation}
To find in full the transformed action with respect to the self propulsion in $d-$dimensions, \Eref{d_dimensional_w_pert}, would be quite the undertaking, however we can elegantly use the isotropy of the system and only consider the $k_d$ component of the transformed position, corresponding to motion in the $\cos\theta_1$ direction. We use the recursive relation of the hyperspherical harmonics with fixed $\mvec=0$ \cite{Wen1985} to write,
    \begin{align}
        \cos{\theta_1} Y_\ell^\nullvec(\Theta)&=\sqrt{\frac{(\ell+1)(\ell+d-2)}{(2\ell+d-2)(2\ell+d)}}Y_{\ell+1}^\nullvec(\Theta) +\sqrt{\frac{\ell(\ell+d-3)}{(2\ell+d-2)(2\ell+d-4)}} Y_{\ell-1}^\nullvec(\Theta) \ ,
    \end{align}
where $Y_{-1}^\nullvec=0$. The perturbative vertices proportional to $k_d$ with fixed $\mvec=\nullvec$ are given by
\begin{subequations}
\begin{align}
\elabel{d_w_vertex1}
     \velpertXa{\ell',\nullvec}{\ell,\nullvec}\triangleq& -\imag
 v k_d \sqrt{\frac{(\ell+1)(\ell+d-2)}{(2\ell+d-2)(2\ell+d)}}\delta_{\ell',\ell+1} \ ,\\
 \elabel{d_w_vertex2}
     \velpertXb{\ell',\nullvec}{\ell,\nullvec}\triangleq& - \imag 
 v k_d \sqrt{\frac{\ell(\ell+d-3)}{(2\ell+d-2)(2\ell+d-4)}}\delta_{\ell',\ell-1} \ ,
\end{align}
\end{subequations}

\subsection{Mean squared displacement}

 We can write the MSD in terms of $k_d$ as 
\begin{equation}
\elabel{d_msd_k_space}
    \ave{|\xvec|^2(t)}=-d\int \dbar{\omega}\exp{-\imag 
    \omega t}\partial^2_{k_d}\bigg|_{\kvec=\nullvec}\ave{\chi^\nullvec_0(\kvec,\omega)\tilde{\chi}^\nullvec_0(-\kvec,-\omega)} \ .
\end{equation}

As in the $d=2$ case, to calculate the MSD  \Eref{d_msd_k_space}, the second order $k_d$ derivatives limit the number of self propulsion vertices, \Erefs{d_w_vertex1} and \eref{d_w_vertex2}, that can be used to construct contributing diagrams. Further, the tumble vertex, \Eref{d_tumble_vertex}, can be arbitrarily many times as in \Eref{TnDef}. We find the MSD to be 
\begin{equation}
\elabel{eq:dMSD}
    \ave{|\xvec|^2(t)}=2D d t+ \frac{2 v^2}{\Lambda^2}(\exp{-\Lambda t}-1+\Lambda t) \ ,
\end{equation}
where the effective persistence is 
\begin{equation}
    \Lambda=\alpha\left(1-\frac{S_d (d-2)}{d}\Pi^{(d)}_1\right) \ .
\end{equation}
We can immediately see that (in having assumed a non-chiral tumble distribution), one can map directly to the two-dimensional MSD, \Eref{MSD_2d}, with $\Pi_1 \to \dfrac{S_d(d-2)}{d} \Pi^{(d)}_1$ (noting that the $d-2$ cancels with a $1/(d-2)$ factor in the Gegenbauer coefficient $\Pi^{(d)}_1$). We further see that once again, as in \Eref{deff}, we reach diffusive behaviour at long-times with
\begin{align}
    D^{(d)}_\textrm{eff} = D + \frac{v^2}{d\Lambda} \ .
\end{align}

\subsection{A particular example: von Mises-Fisher distributions}

To investigate how the orientation distribution influences motion across different dimensions, we consider the von Mises-Fisher distribution and compare our analytical results with numerical data. Using the von Mises-Fisher distribution \cite{Fisher1953, Khatri1977} is defined as
\begin{equation}
\elabel{von_Mises_Fisher}
    \Pi(\dvec\cdot\dvec')=\norm \exp{\kappa \dvec\cdot\dvec'} \ ,
\end{equation}
where  $\norm$ is the normalization, and $\kappa\geq0$ is the concentration parameter. When $\kappa=0$,  the distribution in \Eref{von_Mises_Fisher} is uniform. Substituting the von Mises-Fisher distribution \Eref{von_Mises_Fisher} into the corresponding Gegenbauer transformation, the effective persistence becomes
\begin{equation}
    \effetumbleRate=\alpha\left(1-\frac{I_{d/2}(\kappa)}{I_{d/2-1}(\kappa)}\right) \ ,
\end{equation}
where $I_{d/2}(\kappa)$ is the modified Bessel function of the first kind. When $\kappa=0$ for $d\geq 2$,  the second term in the bracket vanishes, and the effective persistence reduces to $\tumbleRate$, as expected.

\begin{figure}[t!]
    \centering
    \includegraphics[width=0.6\linewidth]{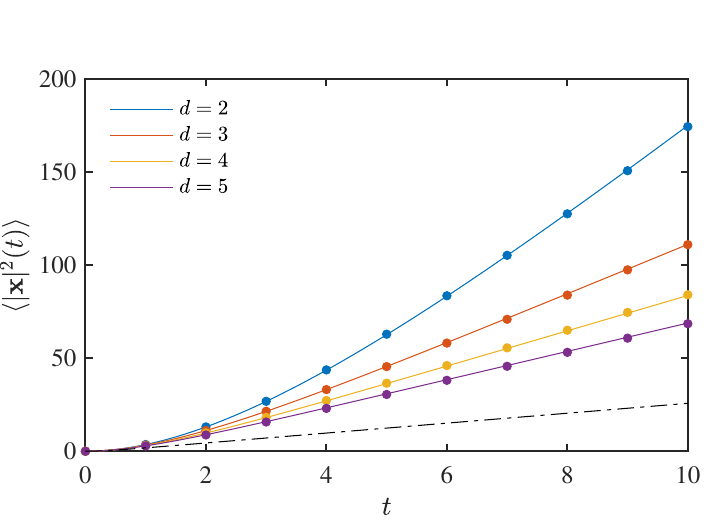}
    \caption{\textit{Mean squared displacement for higher dimensions} -- the MSD of a $d$-dimensional run-and-tumble particle with a director transition distribution given by the von Mises-Fisher distribution \Eref{von_Mises_Fisher}, plotted for $d\in\{2,3,4,5\}$. The black dotted line is the MSD when $\kappa=0$, \ie the director transition is uniform. We use parameters $D=0$; $v=2$; $\alpha=3$; $\kappa=5$. The symbols indicate simulation data and solid lines indicate analytic results, \Eref{eq:dMSD}.}
    \label{fig:datad}
\end{figure}

A comparison between the analytical results and simulation data is shown in Fig.~\ref{fig:datad}. The random number generators used for the 
$d$-dimensional von Mises-Fisher distribution are described in \cite{Pinzon2023}.

\bibliographystyle{iopart-num}
\bibliography{references}

\end{document}